\documentclass[12pt]{elsarticle}
\usepackage{rotating}
\usepackage{mathtools}
\usepackage{comment}
\usepackage{xcolor}
\usepackage{graphicx}
\usepackage{subfig}
\usepackage{amsmath}
\usepackage{amssymb}
\usepackage{enumerate}
\usepackage{natbib}
\usepackage{url} % not crucial - just used below for the URL 
\usepackage{algorithm}
\usepackage{algpseudocode}
\usepackage{booktabs}
\usepackage{caption}
\usepackage[colorlinks=true]{hyperref}
\newtheorem{theorem}{Theorem}
\newtheorem{proposition}{Proposition}
\newcommand{\blind}{0}

\date{}

\begin{document}
	
	\begin{frontmatter}

	\if0\blind
	{
		\title{ Efficient Estimation of COM-Poisson Regression and Additive Model} 
	%%%%%
    \author[rvt]{Suneel Babu Chatla\corref{cor1}}
    \ead{suneel.chatla@iss.nthu.edu.tw}
    \author[rvt]{Galit Shmueli}
    \ead{galit.shmueli@iss.nthu.edu.tw}
    
    \address[rvt]{Institute of Service Science , National Tsing Hua University, Hsinchu 30013, Taiwan}
    \cortext[cor1]{Corresponding author}
	} \fi
	
	\if1\blind
	{
		\title{ Efficient Estimation of COM-Poisson  Regression and Additive Model}
		
		\medskip
  
	} \fi

\begin{abstract}
		The Conway-Maxwell-Poisson (CMP) or COM-Poison  regression is a popular model for count data due to its ability to capture both under dispersion and over dispersion. However, CMP regression is limited when dealing with complex nonlinear relationships. With today's wide availability of count data, especially due to the growing collection of data on human and social behavior, there is need for count data models that can capture complex nonlinear relationships. One useful approach is additive models; but, there has been no additive model implementation for the CMP distribution. To fill this void, we first propose a flexible estimation framework for CMP regression based on iterative reweighed least squares (IRLS) and then extend this model to allow for additive components using a penalized splines approach. Because the CMP distribution belongs to the exponential family, convergence of IRLS is guaranteed under some regularity conditions. Further, it is also known that  IRLS provides smaller standard errors compared to gradient-based methods. We illustrate the usefulness of this approach through extensive simulation studies and using real data from a bike sharing system in Washington, DC.
\end{abstract}
	
	
\end{frontmatter}	

	\section{Introduction}
 
    Count data have become popular dependent variables in studies in various areas, especially due to the growing availability of data on human and social behavior. Examples include the number of crimes  in each neighborhood, number of accidents at an intersection, number of Facebook comments, ridership in bike sharing programs, etc. The wide availability of count data and the need for modeling such data as a function of other factors to establish causal relationships or to quantify correlated relationships has led to the widespread use of count data models. 
    
    The most commonly used regression models for cross-sectional count data are Poisson regression and Negative-Binomial regression. In addition, the Conway-Maxwell-Poisson (CMP) distribution (also known as the COM-Poisson distribution) has gained increasing popularity for its flexibility and ability to handle both over and under dispersed data. 
Revived by \citet{shm05}, the CMP distribution is a two-parameter generalization of the Poisson, Bernoulli, and Geometric distributions. Suppose $Y$ is a random variable that follows a CMP distribution, then the probability mass function (p.m.f.) for $Y\in \{0,1,2,\ldots \} $ is defined as
	\begin{align*}
	P(Y = y)  = \frac{\lambda^y}{(y!)^\nu \zeta (\lambda, \nu)},  \quad \text{where} \quad \zeta(\lambda, \nu) = \sum_{s=0}^{\infty} \frac{\lambda^s}{(s!)^\nu}
	\end{align*}
	for the parameters $\lambda,\nu >0$ and $0 < \lambda <1 , \nu=0 $.
	
	The CMP distribution includes three well-known distributions as special cases: Poisson $(\nu=1)$, Geometric $(\nu=0, \lambda < 1)$, and Bernoulli $(\nu \rightarrow \infty$ with probability $\frac{\lambda}{1+\lambda})$. Due to the additional parameter $\nu$, the CMP distribution is flexible enough to handle both over dispersion ($\nu < 1$) and under dispersion ($\nu > 1$)  which are common in count data \citep{sel08}. For more details on the distributional properties please refer to \cite{dal16}.  
	
	One of the major limitations of  the CMP distribution is that the normalizing constant $\zeta(\lambda,\nu)$, which is an infinite series, does not have a closed form representation, and therefore there is no closed form representation available for the mean. This makes it difficult to model the mean directly as a function of covariates, as in standard models such as Poisson and Logistic regression. However,  the CMP distribution belongs to the exponential family and thus has the properties and advantages of that family. Defining $\boldsymbol{\theta}=(\lambda,\nu)$,  the CMP likelihood has the following form of an exponential family \cite{leh06}:
	\begin{align*}
	P_Y(y|\boldsymbol{\theta}) &= h(y) \exp\bigg(\sum_{i=1}^2\eta_i(\boldsymbol{\theta}) T_i(y) - A(\boldsymbol{\theta}) \bigg),
	\end{align*}
	where the natural parameters are $\eta_1(\boldsymbol{\theta})=\ln \lambda$ and $\eta_2(\boldsymbol{\theta})= -\nu$ with  corresponding  statistics $T_1(y)=y, T_2(y)=\ln(y!)$ and $A(\boldsymbol{\theta})=\ln \zeta(\lambda,\nu), h(y)=1$, as mentioned in \cite{shm05}.

 Although CMP regression is flexible in terms of handling both over and under dispersion, it is sometimes too restrictive for modeling nonlinear relationships or time series data.
	At the same time, additive models are widely  used for modeling nonlinear relationships such as time series \citep{dom02,sti03}. Additive models have the advantage of being parsimonious while at the same time providing more flexibility to capture complicated relationships. Currently, there exists no additive model implementation for the CMP regression. Motivated by the need for flexible count data regression models for applications such as bike sharing, which can assist service providers in better management of their resources, we develop an additive model for CMP regression. Existing additive model implementations are heavily dependent upon the iterative reweighted least squares (IRLS) estimation framework, which currently does not exist for CMP regression. In this study, we propose and implement an IRLS estimation framework for CMP regression and then extend that to additive models.

	 The outline of this paper is as follows: In Section \ref{sec:cmp-reg}, we describe the CMP regression and the problems associated with IRLS implementation. In Section \ref{sec-IRLS}, we develop an IRLS framework for estimating a CMP regression by providing theory and the pseudo algorithm. We evaluate our proposed IRLS methodology with the existing methods using an extensive simulation study in Section \ref{sec-simulation-irls}. In Section \ref{sec-GAM}, we use the IRLS framework to develop an additive model for the CMP distribution, and again evaluate its performance using a simulation  in  Section \ref{sec-simulation-gam}. In Section \ref{sec-bikeshar}, we use our proposed additive model to draw valuable insights from a bike sharing application. Section \ref{sec-conclude} presents conclusions and future directions.
	
	\section{CMP Regression}\label{sec:cmp-reg}
	Assume that we have a random sample of $n$ observations $\{y_i,\boldsymbol{x}_i^T,\boldsymbol{z}_i^T \}_{i=1}^n $, where $\boldsymbol{x}_i^T=[1, x_{i1}, \cdots, x_{ip}]$ and $\boldsymbol{z}_i^T=[1, z_{i1}, \cdots, z_{iq}]$. In matrix notation, let $Y=[y_1, \ldots, y_n]^T$, $X = [\boldsymbol{x}_1,\ldots,\boldsymbol{x}_n]^T$ and $Z=[\boldsymbol{z}_1,\ldots,\boldsymbol{z}_n]^T$ with the parameter vectors $\boldsymbol{\lambda}=(\lambda_1,\ldots,\lambda_n)^T$, $\boldsymbol{\nu}=(\nu_1,\ldots,\nu_n)^T$ and $\boldsymbol{\zeta}=(\zeta_1,\ldots,\zeta_n)^T$. We also denote mean and variance functions as $E[\cdot], V[\cdot]$ respectively. 
	
	When needed, we use the vector notation. With a slight abuse of notation, we extend the operations on scalars to operations on vectors. For example, we write $\ln (Y!)=(\ln(y_1!),\ldots, \ln(y_n!))^T$, $\ln(\boldsymbol{\lambda})=(\ln(\lambda_1),\ldots,\ln(\lambda_n))^T$ and $\frac{\partial \ln \boldsymbol{\zeta}} {\partial \ln \boldsymbol{\lambda}}=(\frac{\partial \ln \zeta_1} {\partial \ln \lambda_1},\ldots,\frac{\partial \ln \zeta_n} {\partial \ln \lambda_n})^T$. Unless otherwise stated, any operation on a vector simply denotes an extension of that operation to each component of that vector.
	
	The CMP regression can be formulated as 
	\begin{align}
	\ln(\boldsymbol{\lambda}) &= X\boldsymbol{\beta} \\
	\ln(\boldsymbol{\nu}) &= Z\boldsymbol{\gamma} 
	\label{eqn:z}
	\end{align}
	where $\boldsymbol{\beta} \in \mathbb{R}^{p+1}, \boldsymbol{\gamma} \in \mathbb{R}^{q+1}$.

	The  log link is used for the  $\boldsymbol{\lambda}$ model. As mentioned in \citet{sel08}, this choice of log link is useful for two reasons. First, it coincides with the link function in two well-known cases: in Poisson regression, it reduces to $E[y_i] = \lambda_i$; in logistic regression, where $p_i =\frac{\lambda_i}{1+\lambda_i}$, it
reduces to $logit(p_i) = \ln \lambda_i$. The second advantage of using a log link function is that it leads to elegant estimation, inference, and diagnostics. At the same time, we deliberately consider a log link for the $\boldsymbol{\nu}$ model, although the canonical link is identity, to restrict model predictions to the range $(0, \infty)$. This is important because while $\boldsymbol{\gamma} $ is unconstrained, $\nu_i$ $(i=1,\ldots,n)$ can only take positive values and we cannot use the identity link between $\boldsymbol{\nu}$ and $\boldsymbol{\gamma}$.
	
	In applications, it is common to treat $\boldsymbol{\nu}$ as nuisance parameter. For this reason, usually the data matrix $Z$ contains only the intercept. Yet, since the $\boldsymbol{\nu}$ parameter models the dispersion, it is always better to include covariates that can potentially control for it \citep{sel13}. In theory, one could use the same predictors for modeling both parameters. However, in practice, to avoid  collinearity issues, it is better to have at least one different covariate in either the $\ln(\boldsymbol{\lambda})$ or the $\ln(\boldsymbol{\nu})$ model. 
	
	Using this model formulation, the log likelihood for the $i^{th}$ observation can be written as 
	\begin{align*}
	\ell_i(y_i, \boldsymbol{\beta}, \boldsymbol{\gamma} ) &=   y_i \boldsymbol{x}_i^T\boldsymbol{\beta}-\ln(y_i!) \exp\{\boldsymbol{z}_i^T\boldsymbol{\gamma}\}- \ln  \zeta_i (\exp\{\boldsymbol{x}_i^T\boldsymbol{\beta}\},\exp\{\boldsymbol{z}_i^T\boldsymbol{\gamma}\}) ,
	\end{align*}
	which yields the following score equations:
	\begin{align}
	\begin{split}
	 \frac{\partial \ell_i}{\partial \boldsymbol{\beta}^T} &= \boldsymbol{x}_i\left(y_i-\frac{\partial \ln \zeta_i} {\partial \ln \lambda_i}\right) =\boldsymbol{x}_i(y_i-E[y_i]),\\
	\frac{\partial \ell_i}{\partial \boldsymbol{\gamma}^T} &= \boldsymbol{z}_i \bigg[\left(-\ln(y_i!)-\frac{\partial \ln \zeta_i} {\partial  \nu_i}\right)\nu_i \bigg]=\boldsymbol{z}_i \bigg[\left(-\ln(y_i!)+E[\ln(y_i!)]\right) \bigg]\nu_i.
	\end{split}
	\label{eqn:score}
	\end{align}
	
	Since the derivatives of $\zeta_i$ $(i=1,\ldots,n)$  do not have closed form representations, the score equations in (\ref{eqn:score}) cannot be solved as in standard generalized linear models (GLM) such as Poisson. For this reason, the existing implementations of CMP regression either use numerical gradient-based methods or Markov Chain Monte Carlo (MCMC), but do not use IRLS, which is the workhorse routine for estimation of all the standard GLMs. Although gradient-based methods have a faster convergence rate than IRLS, they are not efficient because they use the observed information matrix, and are not robust to outliers. In contrast,  IRLS  is more efficient and robust but is  slower than gradient-based methods \citep{gre84}. 
	
	Another advantage of the IRLS algorithm is that modeling extensions such as additive models and lasso can be implemented easily \citep{yee07}. To the best of our knowledge, there is no implementation of an IRLS algorithm for CMP regression. While \citet{sel08} briefly outlined the IRLS algorithm, they did not implement it. Their approach is based on solving the following weighted least squares (WLS) equation at the $m^{th}$ iteration (in matrix notation):
	\begin{align*}
	\begin{bmatrix} X^T \\ (g(Y)*Z)^T\end{bmatrix} W \begin{bmatrix} X & (g(Y)*Z)\end{bmatrix} \begin{bmatrix} \boldsymbol{\beta}^{(m)} \\ \boldsymbol{\gamma}^{(m)}\end{bmatrix} &= \begin{bmatrix} X^T \\ (g(Y)*Z)^T\end{bmatrix} W \quad T
	\end{align*}
	where $W=\text{diag}(V[y_1],\ldots,V[y_n])$, $g(Y)=(g(y_1),\ldots,g(y_n))^T$ with $ g(y_i)=\frac{-\ln(y_i!)+E[\ln(y_i!)]}{y_i-E[y_i]}\nu_i$ and the adjusted dependent variable is $T=(t_1,\ldots,t_n)^T$ with
	\begin{align*}
	t_i &= \boldsymbol{x}_i^T\boldsymbol{\beta}^{(m-1)}+g(y_i)\boldsymbol{z}_i^T\boldsymbol{\gamma}^{(m-1)}+\frac{y_i-E[y_i]}{V[y_i]}.
	\end{align*}

	\citet{sel08} used only an intercept in the $Z$ matrix and did not use a log link function for $\boldsymbol{\nu}$. Here we generalize their approach using a log link function for $\boldsymbol{\nu}$ and not restricting $Z$ to have only an intercept. While the approach looks reasonable, it has the following two drawbacks:
	\begin{enumerate}
		\item  The formulation by \citet{sel08} does not use the expected information matrix. For example, based on their WLS formulation, the information for the intercept term in the $\ln(\boldsymbol{\nu})$ model can be written %computed 
		as: 
\begin{align}
\begin{split}
       \sum_{i=1}^n  g(y_i)^2 V[y_i] &= \sum_{i=1}^n \bigg[\frac{-\ln(y_i!)+E[\ln(y_i!)]}{y_i-E[y_i]} \bigg]^2\nu_i^2 V[y_i]  \\ &\neq \sum_{i=1}^n V[\ln(y_i!)]\nu_i^2.
\end{split}
\end{align}
       The value in the right hand side of the inequality  is the expected information for the intercept term in the  $\ln(\boldsymbol{\nu})$ model using the score equations (\ref{eqn:score}). Clearly, there is some discrepancy as the information evaluated from the \citet{sel08} derivation is not equal to the expected information. What we know is that the expected information is efficient and we do not know whether the \citet{sel08} formula achieves the same efficiency (at least asymptotically). The \citet{sel08} formulation matches the expected information only if
       \begin{align*}
           \bigg[\frac{-\ln(y_i!)+E[\ln(y_i!)]}{y_i-E[y_i]} \bigg]^2 &= \frac{E\left[-\ln(y_i!)+E[\ln(y_i!)]\right]^2}{E\left[y_i-E[y_i]\right]^2}.
       \end{align*}
       
       Based on the well known \emph{Cram\`{e}r - Rao inequality} \cite{kee06}, the variance of any unbiased estimator is bounded by the inverse of the expected (Fisher) information. In general, the IRLS should use the expected information.   
		\item The idea of combining both models into a single WLS framework is computationally attractive. However, since both $\boldsymbol{\beta}$ and $\boldsymbol{\gamma}$ are dependent on each other, updating both of them in single model is problematic especially with least squares. When we implemented this approach, most of the time the algorithm remained close to the initial values and sometimes it chose very small values of $\nu_i$ $(i=1,\ldots,n)$ irrespective of the true values. 
	\end{enumerate}
	
	To overcome these limitations, we propose a two step IRLS algorithm with guaranteed convergence that uses the expected information matrix for updates. Our approach also makes it easier to extend the CMP regression for the estimation of additive components.

	\section{IRLS Framework for CMP Regression} \label{sec-IRLS}
    
	To implement the IRLS method, we must first calculate the cumulants $E[y_i]$, $E[\ln(y_i!)]$, $V[y_i]$ and $V[\ln(y_i!)]$ for $i=1,\ldots,n$. 
	\subsection{Calculation of Cumulants}
	The standard way of calculating cumulants is to use the p.m.f. Since the p.m.f. for the CMP distribution involves an infinite series, a simple approach is to truncate the infinite series in such a way that the error is bounded ($\epsilon=10^{-6}$) \cite{shm05}.
	
	Another way of calculating the cumulants is by using the properties of the canonical parameter of the exponential family. The $t^{th}$ cumulants for $y_i$ and $\ln(y_i!)$ for $i=1,\ldots, n$ can be obtained as:
	\begin{align*}
	\kappa_{t}[y_i]  = \frac{\partial^{(t)} \ln \zeta_i (\lambda_i,\nu_i)}{\partial^{(t)} \ln\lambda_i} , \quad
	\kappa_{t}[\ln(y_i!)]  =- \frac{\partial^{(t)} \ln \zeta_i (\lambda_i,\nu_i)}{\partial^{(t)} \nu_i}. 
	\end{align*} 
	%To keep this in mind, t
    There has been some active research trying to approximate the $\zeta_i$ function using a closed form representation. \citet{shm05} provided that for fixed positive integer $\nu_i$ the following asymptotic approximation holds: % for the function:
	\begin{align}
	\zeta_i(\lambda_i, \nu_i) & = \frac{e^{\nu_i \lambda_i^{1/\nu_i}}} {\lambda_i^{\frac{\nu_i-1}{2\nu_i}} (2\pi)^{\frac{\nu_i-1}{2}} \sqrt{\nu_i}}  (1+ \mathcal{O}(\lambda_i^{\frac{-1}{\nu_i}})) .
	\label{eqn:z-asym-earl}
	\end{align}
	Earlier, \citet{olv14} had derived the same leading term in the asymptotic expansion (\ref{eqn:z-asym-earl}) and proved that it is valid for $0 < \nu_i\le 4$. \citet{gil15} built on the work of \cite{olv14} to confirm that Equation (\ref{eqn:z-asym-earl}) holds for all $\nu_i > 0$.
	
	For higher order cumulants ($t>1$) the cumulant generating function  has the following form:
	\begin{align}
	\kappa_t[y_i] = \nu_i \lambda_i^{1/\nu_i} (e^{t/\nu_i}-1).
	\label{eqn:asym-cum}
	\end{align}
	Although this approximation is appealing theoretically, it has limited practical value. To get a better approximation with the formulation in Equation (\ref{eqn:asym-cum}) we should have larger $\lambda_i^{1/\nu_i}$ values, i.e., larger counts.
	
	Recently, \citet{gau16} further improved the asymptotic approximation in Equation (\ref{eqn:z-asym-earl}) by providing lower order terms : % as following:
	\begin{align}
	\zeta_i (\lambda_i,\nu_i) =\frac{e^{\nu_i \lambda_i^{1/\nu_i}}} {\lambda_i^{\frac{\nu_i-1}{2\nu_i}} (2\pi)^{\frac{\nu_i-1}{2}} \sqrt{\nu_i}}  \left(1+c_1(\nu_i\lambda_i^{1/\nu_i})^{-1}+c_2(\nu_i\lambda_i^{1/\nu_i})^{-2}+ \mathcal{O}(\lambda_i^{\frac{-3}{\nu_i}})\right), 
	\label{newz}
	\end{align}
	where $c_1=\frac{\nu_i^2-1}{24}, c_2=\frac{\nu_i^2-1}{48}+\frac{c_1^2}{2}$.  \citet{dal16} derived the leading term in the asymptotic approximation of all cumulants. However,  since \citet{gau16}  provided the expressions for the cumulants for $E[y_i]$ and $V[y_i]$ including the first two correction terms, we use their results to approximate the mean and variance. Define $\alpha_i=\lambda_i^{1/\nu_i}$, then
	\begin{align}
	   E[y_i] &= \alpha_i-\frac{\nu_i-1}{2\nu_i}-\frac{\nu_i^2-1}{24\nu_i^2}\alpha_i^{-1}-\frac{\nu_i^2-1}{24\nu_i^3}\alpha_i^{-2}+\mathcal{O}(\alpha_i^{-3}),\\
	V[y_i] &= \frac{\alpha_i}{\nu_i}+  \frac{\nu_i^2-1}{24\nu_i^3}\alpha_i^{-1}+\frac{\nu_i^2-1}{12\nu_i^4}\alpha_i^{-2}+\mathcal{O}(\alpha_i^{-3}).
	\end{align}
	
	Since we also need the first two cumulants for $\ln(y_i!)$, we used the asymptotic expression in (\ref{newz}) to calculate both $E[\ln(y_i!)]$ and $V[\ln(y_i!)]$:
	\begin{align}
	\begin{split}
	E[\ln(y_i!)] &=\alpha_i \left(\frac{\ln\lambda_i}{\nu_i}-1 \right) +\frac{\ln\lambda_i}{2\nu_i^2}+\frac{1}{2\nu_i}+\frac{\ln2\pi}{2} \\& -\frac{\alpha_i^{-1}}{24}\left(1+\frac{1}{\nu_i^2}+\frac{\ln\lambda_i}{\nu_i}-\frac{\ln\lambda_i}{\nu_i^3} \right)   \\&-\frac{\alpha_i^{-2}}{24}\left(\frac{1}{\nu_i^3}+\frac{\ln\lambda_i}{\nu_i^2}-\frac{\ln\lambda_i}{\nu_i^4} \right)+  \mathcal{O}(\alpha_i^{-3}),  
	\end{split}
	\\
	%\end{align}
	%\begin{align}
	\begin{split}
	    V[\ln(y_i!)] &=\alpha_i\frac{(\ln\lambda_i)^2}{\nu_i^3}+\frac{\ln\lambda_i}{\nu_i^3}+\frac{1}{2\nu_i^2}\\&
	    + \frac{\alpha_i^{-1}}{24\nu_i^5}\left( -2\nu_i^2+4\nu_i\ln \lambda_i+(-1+\nu_i^2)(\ln \lambda_i)^2 \right) \\&
	    +\frac{\alpha_i^{-2}}{24\nu_i^6} \left(-3\nu_i^2-2\nu_i(-3+\nu_i^2)\ln\lambda_i+2(-1+\nu_i^2)(\ln\lambda_i)^2 \right)
	    +\mathcal{O}(\alpha_i^{-3}).
	\end{split}
	\end{align}

	\begin{figure}[t]
		\centering
		\includegraphics[scale=0.8]{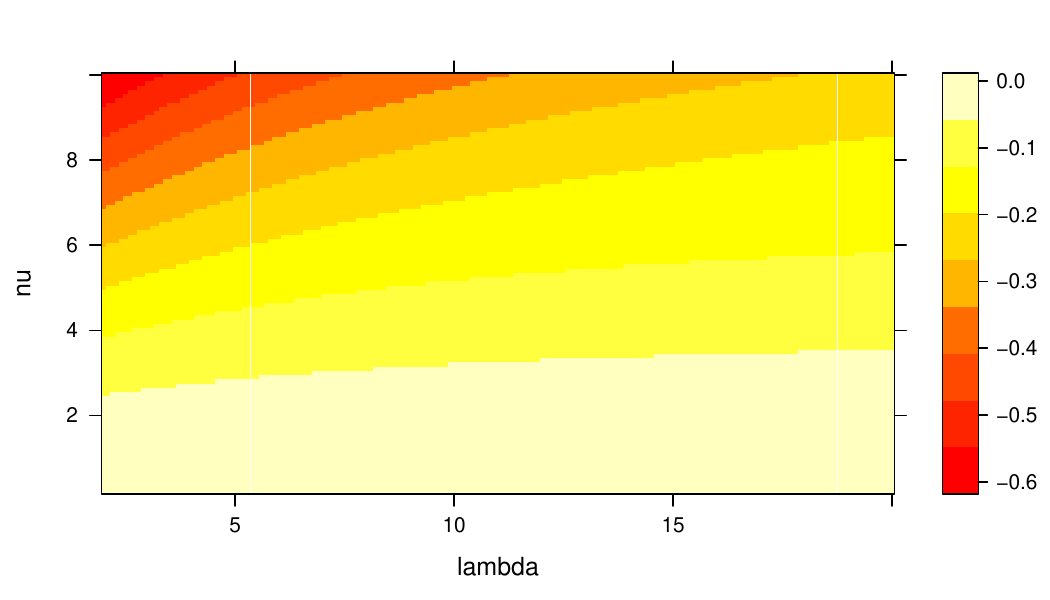}
		\caption{The differences between $\ln \zeta_i $ calculated using the truncated infinite series and the asymptotic expression in (\ref{newz}) for the parameter range: $2 \leq\lambda_i \leq 20$, $0.2 \leq \nu_i \leq 10$.}
		\label{z-approx}
	\end{figure}
	
	While \citet{gau16} showed that the asymptotic approximation is much better after including two more terms, it is still not close to the true value at least for the parameter range $0<\lambda_i<2$. Hence we consider the remaining range to check whether the asymptotic approximation is accurate or not. To illustrate this, we computed the $\ln \zeta_i$ function for the parameter range $2\le \lambda_i\le 20$, $0.2 \le \nu_i \le 10$ using both the truncated infinite series (bounding error=$10^{-6}$) and the asymptotic expression in (\ref{newz}). We plotted the differences between the two sets of values in Figure \ref{z-approx}. It can be observed that the new approximation is reasonably good  when $\lambda_i\geq 2$ and $\nu_i \leq 1$, while for higher values of $\nu_i$ $(>2)$ the asymptotic approximation tends to over estimate the true value. For this reason we use the cumulants derived from the asymptotic expression only when $\lambda_i\geq 2$, $ \nu_i \leq 1$, while for other values  we use the p.m.f. to calculate cumulants recursively with some bounding error. Although the approximation works for a limited range, this is very helpful because the asymptotic series converges very slowly when $\nu_i <1$ and this approximation % definitely 
    eases the computational burden significantly. 

Similarly, for  $\ln(y_i!)$ the values are computed recursively until $y_i <254$ and after that  Stirling's approximation is used as it is reasonably close \citep{abr66}.

	\subsection{Two step method}
	
	Let us define $\boldsymbol{u}_{(p+1)\times 1}=\sum_{i=1}^n \frac{\partial \ell_i}{\partial \boldsymbol{\beta}^T}$ and $\boldsymbol{v}_{(q+1)\times 1}=\sum_{i=1}^n \frac{\partial \ell_i}{\partial \boldsymbol{\gamma}^T}
	$. From Equation (\ref{eqn:score}), the full information matrix $I$ can be written as %obtained as follows
	\begin{align*}
	I_{(p+1)\times (q+1) } = E \Bigg[\begin{pmatrix}\boldsymbol{u} \\\boldsymbol{v}\end{pmatrix}(\boldsymbol{u}^T \boldsymbol{v}^T) \Bigg] 
	= \begin{bmatrix} E[\boldsymbol{u}\boldsymbol{u}^T] & E[\boldsymbol{u}\boldsymbol{v}^T] \\
	E[\boldsymbol{v}\boldsymbol{u}^T] & E[\boldsymbol{v}\boldsymbol{v}^T]  
	\end{bmatrix} 
	= \begin{bmatrix} I_{11} & I_{12} \\
	I_{21} & I_{22}  
	\end{bmatrix},
	\end{align*}
	with 
	\begin{align*}
	E[\boldsymbol{u}\boldsymbol{u}^T] &= I_{11}= X^T\Sigma_YX ,\\
	E[\boldsymbol{u}\boldsymbol{v}^T] &= I_{12}=I^T_{21}=-\boldsymbol{\nu}* X^T\Sigma_{Y,\ln (Y!)} Z,\\
	E[\boldsymbol{v}\boldsymbol{v}^T] &= I_{22}=\boldsymbol{\nu}^2* Z^T\Sigma_{\ln (Y!)}Z,
	\end{align*}
    where $*$ denotes element-wise multiplication, $\Sigma_Y = \text{diag}(V[y_1],\ldots,V[y_n])$, $\Sigma_{Y,\ln (Y!)}= \left[\text{cov}(y_i,\ln(y_j!))\right]_{1\le i,j \le n}$,  and  $\Sigma_{\ln (Y!)}=\text{diag}(V[\ln(y_1!)],\ldots,V[\ln(y_n!)])$.
	
	Using the information matrix $I$,  the IRLS algorithm has the following form for the $m^{th}$ update:
	\begin{align*}
	\begin{bmatrix}
	\boldsymbol{\beta} \\ \boldsymbol{\gamma}
	\end{bmatrix}^{(m)} =  \begin{bmatrix}
	\boldsymbol{\beta} \\ \boldsymbol{\gamma}
	\end{bmatrix}^{(m-1)} + I^{-1} \begin{bmatrix}
	\boldsymbol{u} \\ \boldsymbol{v}
	\end{bmatrix},
	\end{align*}
	which implies the following equations:
	\begin{align}
	\begin{split}
	\label{eqn:wls1}
	 & X^T\Sigma_YX\boldsymbol{\beta}^{(m)}-X^T\Sigma_{Y,\ln (Y!)}\boldsymbol{\nu}* Z\boldsymbol{\gamma}^{(m)} = \\   & X^T\Sigma_YX\boldsymbol{\beta}^{(m-1)}-  
 X^T\Sigma_{Y,\ln (Y!)}\boldsymbol{\nu}* Z\boldsymbol{\gamma}^{(m-1)} 
	+ X^T(Y-E[Y])
	\end{split}
	\end{align}
	and
	\begin{align}
	\begin{split}
	\label{eqn:wls2}
	&-\boldsymbol{\nu}* Z^T\Sigma_{Y,\ln (Y!)}X\boldsymbol{\beta}^{(m)}+\boldsymbol{\nu}^2* Z^T\Sigma_{\ln (Y!)}Z\boldsymbol{\gamma}^{(m)} =  \\ &-\boldsymbol{\nu} Z^T\Sigma_{Y,\ln (Y!)} X\boldsymbol{\beta}^{(m-1)}+ \boldsymbol{\nu}^2* Z^T\Sigma_{\ln(Y!)}Z\boldsymbol{\gamma}^{(m-1)} \\ &
	+ \boldsymbol{\nu}*Z^T(-\ln (Y!)+E[\ln (Y!)]). 
	\end{split}
	\end{align}

	Each of the  two equations in (\ref{eqn:wls1}) and (\ref{eqn:wls2}) are complicated and contain updates for both parameters $\boldsymbol{\beta}$ and $\boldsymbol{\gamma}$. However, if we fix one parameter in each equation, a nice closed form expression appears for the updates. When we fix $\boldsymbol{\gamma}$ in equation (\ref{eqn:wls1}) the equation reduces to 
	\begin{align}
	\label{eq:mwls1}
	X^T\Sigma_YX\boldsymbol{\beta}^{(m)}&=X^T\Sigma_YX\boldsymbol{\beta}^{(m-1)}+X^T(Y-E[Y]).
	\end{align}
	This equation is nothing but WLS of $X$ on $Y$ with weights $\Sigma_Y$. Similarly, if we fix $\boldsymbol{\beta}$ in equation (\ref{eqn:wls2}) then the equation reduces to
	\begin{align}
	\label{eq:mwls2}
	\boldsymbol{\nu}^2* Z^T\Sigma_{\ln(Y!)}Z\boldsymbol{\gamma}^{(m)}&=\boldsymbol{\nu}^2*  Z^T\Sigma_{\ln (Y!)}Z\boldsymbol{\gamma}^{(m-1)}+\boldsymbol{\nu}* Z^T(-\ln (Y!)+E[\ln (Y!)]).
	\end{align}
	Again this is a WLS of $\boldsymbol{\nu}*Z$ on $\ln(Y!)$ with weights $\Sigma_{\ln(Y!)}$.
	
	The two update equations (\ref{eq:mwls1}) and (\ref{eq:mwls2}) are elegant and can be easily estimated with WLS methods. This approach is not only convenient for estimation but also  helpful for generalizing to other modeling extensions such as additive models and the lasso.

\subsection{Proof of Convergence of the Two Step Method}
	To prove the convergence properties of our proposed two step algorithm, we start with the following assumptions. Consider the parameter space $\Theta \in (0,\infty) \times (0,\infty)$, and the likelihood function $L$.
	
	\textit{(A1).} Let $\boldsymbol{\hat{\theta}_0}=(\boldsymbol{\hat{\lambda}_0},\boldsymbol{\hat{\nu}_0})^T \in \Theta$ be a starting value, such that $D_0=\{\boldsymbol{\theta}=(\boldsymbol{\lambda},\boldsymbol{\nu})^T \in \Theta \vert L(\boldsymbol{\theta}) \ge L(\boldsymbol{\hat{\theta}_0})\}$ is compact.
	
	\textit{(A2).} The function $L$ is uniquely maximized over $D_0$ for $\boldsymbol{\theta}=\boldsymbol{\hat{\theta}}$.
	
	\textit{(A3).} Suppose that we have given parameter functions 
	$\psi_i:D_0 \to \Theta_i \quad (i=1,2)$
	and let $M_i(\boldsymbol{\theta}), \boldsymbol{\theta} \in D_0$ be the corresponding sections:
	$M_i(\boldsymbol{\theta})=\{\boldsymbol{\eta} \in D_0 \vert \psi_i(\boldsymbol{\eta})=\psi_i(\boldsymbol{\theta})\} \quad
	(i=1,2)$. 
	Then we assume that, for $i=1,2$ and $\boldsymbol{\theta} \in D_0$, $L$ is maximized uniquely  by $T_i(\boldsymbol{\theta})$	on the section $M_i(\boldsymbol{\theta})$ and that $T_i(\boldsymbol{\theta})$ is continuous on $D_0$.
	
	\textit{(A4).} The point of global maximum $\boldsymbol{\hat{\theta}}$ is uniquely determined by the condition that it is the partial maximum along each section $M_i(\boldsymbol{\theta})$. In other words,
    \begin{align*}
    \sup_{\boldsymbol{\eta} \in M_i(\boldsymbol{\theta})} L(\boldsymbol{\eta}) &= L(\boldsymbol{\theta}), \quad i=1,2
    \end{align*}
	implies $\boldsymbol{\theta} = \boldsymbol{\hat{\theta}}$, or equivalently, $T_i(\boldsymbol{\theta})= \boldsymbol{\theta}$ implies  $\boldsymbol{\theta} = \boldsymbol{\hat{\theta}}$.
	
	Assumptions A1 and A2 are based on the fact that the CMP distribution is unimodal and it has a log-concave p.m.f. \citep{Gup14,ste85}. The remaining assumptions A3 and A4  follow from the properties of exponential family distributions. It is well known that the marginal distributions in a regular $k$-variate exponential family  also belong to an exponential family \citep{kee06,leh06}. It means that  for a distribution that belongs to an exponential family like the CMP distribution, the estimates obtained from maximizing  the marginal likelihood are the same as the estimates obtained from maximizing the full likelihood. 

\begin{theorem}
Under assumptions A1-A4, the two step IRLS algorithm 
\begin{align*}
\boldsymbol{\hat{\theta}_{n+1}} &=T_1(T_2(\boldsymbol{\hat{\theta}_n}))
\end{align*}
converges to $\hat{\boldsymbol{\theta}}$ for any starting value in $D_0$.
\end{theorem}
	The proof is similar to \citet{jen91}. The authors showed that under the above assumptions any partial maximization algorithm converges to the true value for a given starting value.

	% Pseudo Algorithm
	\begin{algorithm}
		\caption{IRLS Framework for CMP distribution}
		\begin{algorithmic}[1]
			 \State Set initial values for $\nu_i^{(0)}$ and $\lambda_i^{(0)}=(y_i+0.1)^{\nu_i^{(0)}}$ for $i=1,\ldots,n$.
			 \State Compute $\eta_{i1}^{(0)}=\ln(\lambda_i^{(0)})$ and $\eta_{i2}^{(0)}=\ln(\nu_i^{(0)})$ for $i=1,\ldots,n$.
			 \State Compute $D^{(0)}(\boldsymbol{\lambda}^{(0)},\boldsymbol{\nu}^{(0)})=-2\sum_{i=1}^n\ell(\lambda_i^{(0)},\nu_i^{(0)})$.\\
			 Compute $E[y_i]^{(0)}$ and $V[y_i]^{(0)}$ for $i=1,\ldots,n$. 
			 \For{$k$ in 1:maxIter}
			 \State Compute the adjusted dependent variable for each $i=1,\ldots,n$: $t_{i1}^{(k)}=\eta_{i1}^{(k-1)}+\frac{y_i-E[y_i]^{(k-1)}}{V[y_i]^{(k-1)}}$.
			 \State Perform a weighted least squares regression of $T_1^{(k)}=(t_{11},\ldots,t_{n1})^T$ on $X$ with weights $W_1=\text{diag}(V(y_1)^{(k-1)},\ldots,V(y_n)^{(k-1)})$ to obtain $\boldsymbol{\beta}^{(k)}$.
			 \State update $\eta_{i1}^{(k)}=\boldsymbol{x}_i^T\boldsymbol{\beta}^{(k)}$ and $\lambda_i^{(k)}=\exp(\eta_{i1}^{(k)})$ for $i=1,\ldots,n$.
			 \State Compute $E[\ln(y_i!)]^{(k-1)}$ and $V[\ln(y_i!)]^{(k-1)}$ for $i=1,\ldots,n$.
			 \State Compute the adjusted dependent variable: $t_{i2}^{(k)}=\nu_i^{(k-1)} \eta_{i2}^{(k-1)}+\frac{-\ln(y_i!)+E[\ln(y_i!)]^{(k-1)}}{V[\ln(y_i!)]^{(k-1)}}$.
			 \State Perform a weighted least squares regression of $T_2^{(k)}=(t_{12},\ldots,t_{n2})^T$ on $\boldsymbol{\nu}^{(k-1)} *Z$ with weights $W_2=\text{diag}(V[\ln(y_1!)]^{(k-1)},\ldots,V[\ln(y_n!)]^{(k-1)})$ to obtain $\boldsymbol{\gamma}^{(k)}$.
			 \State Update $\eta_{i2}^{(k)}=\boldsymbol{z}_i^T\boldsymbol{\gamma}^{(k)}$ and $\nu_i^{(k)}=\exp(\eta_{i2}^{(k)})$ for each $i=1,\ldots,n$.
			 \State Compute $D^{(k)}(\boldsymbol{\lambda}^{(k)},\boldsymbol{\nu}^{(k)})=-2\sum_{i=1}^n\ell(\lambda_i^{(k)},\nu_i^{(k)})$.
			 \If{$\frac{D^{(k)}-D^{(k-1)}}{D^{(k)}} > 10^{-6}$}
			 \State Initiate step size optimization.
			 \EndIf
			 \If{$|\frac{D^{(k)}-D^{(k-1)}}{D^{(k)}}| < 10^{-6}$}
			 \State Convergence achieved. Break the loop.
			 \Else
			 \State Compute $E[y_i]^{(k)}$ and $V[y_i]^{(k)}$ for each $i=1,\ldots,n$. 
			 \EndIf
			 \EndFor
		\end{algorithmic}
		\label{alg:IRLS}
	\end{algorithm}

	\subsection{Practical issues}

	\subsubsection{Initial Values}
	Unlike  other nonlinear optimization algorithms, IRLS does not require initial values for the parameters $\boldsymbol{\beta}$ and $\boldsymbol{\gamma}$ but it does require initial values for $\boldsymbol{\lambda}$ and $\boldsymbol{\nu}$. We can provide suitable initial values based on the approximate method of a moments estimator for $ \lambda_i$ such as $(y_i+0.1)^{\nu_i^{(o)}}$ for $i=1,\ldots,n $. However, we do not have a closed form expression for $\nu_i^{(o)}$. In practice we observed that starting close to zero (e.g., $\nu_i =0.2$ or $0.5$) yields satisfactory results.

	\subsubsection{Stopping Criterion} 
	The standard IRLS algorithm uses the deviance as a  stopping criterion. If the absolute relative change in the deviance is below some tolerance threshold, the algorithm stops. In general, the deviance for the $i^{th}$ observation is defined as:
	\begin{align*}
	D_i=-2(\ell(y_i;\hat{\lambda}_i,\hat{\nu}_i)-\ell(y_i;\hat{\lambda}_{i,sat},\hat{\nu}_{i,sat})).
	\end{align*}
	The estimates for both  $\hat{\lambda}_{i,sat},\hat{\nu}_{i,sat}$ depend on each other and we do not have closed forms especially for the estimate $\hat{\nu}_{i,sat}$. For this reason, we consider only the term $-2 \sum \ell(y_i;\hat{\lambda}_i,\hat{\nu}_i)$ and use it as our stopping criterion. Since the likelihood for the saturated model is constant across all the iterations, ignoring the term $\ell(y_i;\hat{\lambda}_{i,sat},\hat{\nu}_{i,sat})$ does not impact our stopping criterion. In addition, this function is monotonic and if the algorithm is converging, it will decrease with every iteration. % as does the regular deviance in GLMs, if the algorithm is converging. %Indeed, this monotonic property is very helpful monitoring the convergence of our algorithm.
	
	\subsubsection{Step Size}
	It is common for IRLS to exhibit convergence problems \citep{mar11}. To avoid non-convergence issues we used the step-halving approach suggested by \citet{mar11}. The algorithm invokes step-halving either at the boundary or if the deviance is increasing. This step-halving makes sure that the algorithm remains in the interior space which is required for convergence.   
	
	\subsection{Inference}
    \begin{proposition}
   Under regularity conditions \citep{kee06}[p.158], the maximum likelihood estimators $\boldsymbol{\hat{\theta}}= (\boldsymbol{\hat{\beta}} ,\boldsymbol{\hat{\gamma}})^T $  
	are consistent and asymptotically normal: 
	\begin{align*}
	\sqrt{n}(\boldsymbol{\hat{\theta}}-\boldsymbol{\theta}) \xrightarrow{\mathit{d}} \mathcal{N}_{p+q+2}(0,I^{-1}(\boldsymbol{\theta})).
	\end{align*} 
 \end{proposition}
	The proof is an immediate consequence of the result from  \citet{kee06}[p.158].
	Since the algorithm estimates each parameter vector separately while keeping the other parameter vector fixed, it only provides the marginal information for the respective parameters. The conditional information matrices can be straightforwardly obtained by using the matrix operations \cite{lu02} as following:
	\begin{align*}
	\sqrt{n}(\boldsymbol{\hat{\beta}}-\boldsymbol{\beta}) \vert \boldsymbol{\hat{\gamma}} \xrightarrow{\mathit{d}} \mathcal{N}_{p+1}(0,(I_{11}-I_{12}I_{22}^{-1}I_{21})^{-1}) \\
	\sqrt{n}(\boldsymbol{\hat{\gamma}}-\boldsymbol{\gamma})\vert \boldsymbol{\hat{\beta}} \xrightarrow{\mathit{d}} \mathcal{N}_{q+1}(0,(I_{22}-I_{21}I_{11}^{-1}I_{12})^{-1}).
	\end{align*}
	
	We note that the estimates for both $\boldsymbol{\beta}$ and $\boldsymbol{\gamma}$ are not independent and  inferences on one parameter will be influenced by the other estimate. As it was mentioned earlier, for most practical applications, inference on the  parameter $\boldsymbol{\beta}$ is of primary interest and usually the parameter $\boldsymbol{\gamma}$ will be treated as a nuisance parameter. 
	
	\section{Simulation Study for the CMP Regression}\label{sec-simulation-irls}

	We conducted an extensive simulation study to evaluate the performance of our proposed IRLS algorithm in comparison to  existing gradient-based methods for estimating the CMP regression model. 
	
	At present there are two R packages (\emph{CMPRegression} by \cite{sel11}; \emph{CompGLM} by \cite{CompGLM}) for fitting the CMP regression model. Both use general purpose optimization functions to maximize the  likelihood function. While these two R packages are technically the same, they differ in terms of their implementations. From now on, we denote these packages as  $Opt_1$ and $Opt_2$ and our implementation as $IRLS$. While $Opt_1$ does not use a log link for the $\boldsymbol{\nu}$ model, $Opt_2$ does use a log link and allows the user to model the $\boldsymbol{\nu}$ parameter as well. Similarly, while $Opt_1$ only provides the log likelihood to the optimization function, $Opt_2$  also provides the gradients. More importantly, in terms of computational issues, the support functions for $Opt_1$ were implemented in R and  the support functions for $Opt_2$ were implemented in C++ which makes it work much faster.
	
	%%%%
		% y plot
	\begin{figure}[!htbp]		
		\centering
			\includegraphics[scale=0.7]{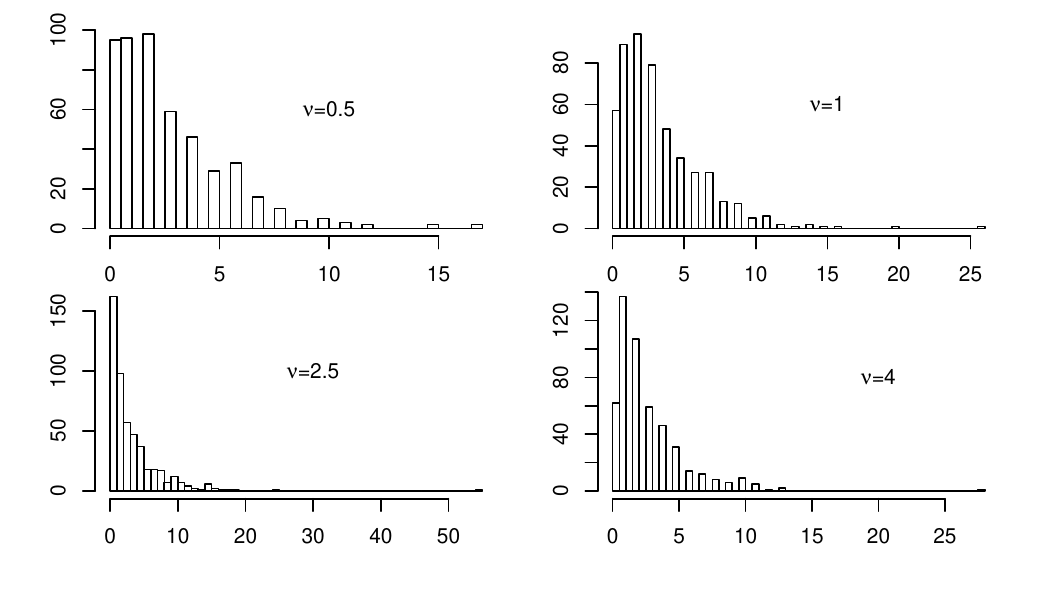}
			\caption{Distribution of the simulated $y$ corresponding to the results in Table \ref{tab:sim-res} ( $\nu=0.5,1,2.5 $ and $4$,  respectively).}
			\label{fig:irls-y-sim}
		\end{figure}

    Both  $Opt_1$ and $Opt_2$ methods have some limitations. One obvious limitation is their inability to handle larger counts. Even with a single large value in the data  both  methods produce errors. Since both methods supply the likelihood to an external optimization routine in R and the source code is not available for the external function, we were not able to identify the reasons as to why these methods fail with larger counts. However, we believe that the problem is related to computing numerical derivatives which is often problematic with the CMP distribution because of the normalizing constant.

	In order to provide a clear comparison of our method with the two aforementioned gradient-based methods, we carefully constructed simulated datasets without any large counts, as shown in Figure \ref{fig:irls-y-sim}, so that none of these methods face any convergence issues. Although the distributions of the dependent variable in Figure \ref{fig:irls-y-sim}, look similar they have different counts (please note the scale on the $x$-axis). After exploring several coefficient values, we reached these datasets for which both $Opt_1$ and $Opt_2$ did not show any problems. For the data simulated from the other sets of coefficient values only $IRLS$ provided results. We considered sample size $n=500$ and chose 4 covariates to be included in the model. The covariates are simulated from normal and uniform distributions and also allow mild correlation between one pair of covariates ($x_1  \sim U(0,1), x_2 \sim N(0,0.5)$, $x_3 \sim N(0,0.1) $ and $x_4 = 0.2 x_3+N(0,0.5)$).
	
	We considered four different values for $\nu$ in order to capture over dispersion $(\nu=0.5)$, equi dispersion $(\nu=1)$ and under dispersion $(\nu=2.5,4)$ scenarios. The true values for the regression coefficients and their estimated values using 20 bootstrap replications are reported in Table \ref{tab:sim-res}. From the results, it can be observed that $IRLS$ performs equally well to the existing gradient-based methods, especially $Opt_1$. While the three methods are indistinguishable for the over dispersion case $(\nu=0.5)$, we  observe that there are some clear discrepancies for the under dispersion case  ($\nu= 2.5$ or $4$).  In particular, $Opt_2$  has some issues when there is under dispersion in the data.

	\begin{table}[!htbp]
		\begin{minipage}[b]{.4\linewidth}
			\centering
			\small
			%\caption{My caption}
			%\label{my-label}
			\begin{tabular}{cllll}
				\toprule
				& \multicolumn{4}{c}{$\nu=0.5$}  \\
				\cmidrule{2-5} 
				& $\theta$ & $\hat{\theta}_{IRLS}$ & $\hat{\theta}_{Opt_1}$ &$\hat{\theta}_{Opt_2}$ \\
				& & (sd) & (sd) &(sd) \\
				\midrule
				$\beta_0$ & \textbf{0.05} & \textbf{0.05} & \textbf{0.04} &\textbf{0.04}    \\
				&  & (0.07) & (0.07) &(0.07)    \\
				$\beta_1$& \textbf{0.5} & \textbf{0.52} &\textbf{0.52}  &\textbf{0.52} \\
				&  & (0.07) &(0.07)  &(0.07) \\
				$\beta_2$& \textbf{-0.5} & \textbf{-0.52} &\textbf{-0.52}  &\textbf{-0.52} \\ 
				&  & (0.06) &(0.06)  &(0.06) \\
				$\beta_3$&  \textbf{0.25}& \textbf{0.31} &\textbf{0.31}  &\textbf{0.31} \\
				&  & (0.21) &(0.21)  &(0.21) \\
				$\beta_4$& \textbf{-0.25} & \textbf{-0.25} &\textbf{-0.25}  &\textbf{-0.25} \\
				&  & (0.06) &(0.06)  &(0.06) \\
				$log(\nu)$& \textbf{-0.69} &\textbf{-0.65}  &\textbf{-0.67}  & \textbf{-0.67}\\
				&  &(0.09)  &(0.09)  & (0.09)\\
				\bottomrule
			\end{tabular}
		\end{minipage}
		\hfill
		\begin{minipage}[b]{0.4\linewidth}
			\small
			\begin{tabular}{ccccc}
				\toprule
				& \multicolumn{4}{c}{$\nu=1$} \\
				\cmidrule{2-5} 
				& $\theta$ & $\hat{\theta}_{IRLS}$ & $\hat{\theta}_{Opt_1}$ &$\hat{\theta}_{Opt_2}$ \\
				& & (sd) & (sd) &(sd) \\
				\midrule
				$\beta_0$ &  \textbf{0.5} & \textbf{0.46} &\textbf{0.50} & \textbf{0.48}    \\
				& & (0.11) & (0.69) & (0.06) \\
				$\beta_1$&\textbf{ 1}  & \textbf{0.99} &\textbf{1.03} & \textbf{1.01} \\
				&  & (0.11) &(0.09) & (0.08) \\
				$\beta_2$&\textbf{-1}  & \textbf{-0.98} & \textbf{-1.03} & \textbf{-1} \\ 
				& & (0.09) & (0.08) & (0.06) \\ 
				$\beta_3$&\textbf{0.5} & \textbf{0.56} & \textbf{0.59}& \textbf{0.58} \\
				& & (0.23) & (0.24)& (0.25) \\
				$\beta_4$&\textbf{-0.5}  &\textbf{-0.49} & \textbf{-0.51}& \textbf{-0.50}\\
				& &(0.08) & (0.08)& (0.07)\\
				$log(\nu)$& \textbf{0}  & \textbf{-0.03}  & \textbf{0.02} &\textbf{0}\\
				&  & (0.10)  & (0.06) &(0)\\
				\bottomrule
			\end{tabular}
		\end{minipage}
		\hfill
		\begin{minipage}[b]{0.4\linewidth}
			\small
			\begin{tabular}{ccccc}
				\toprule
				& \multicolumn{4}{c}{$\nu=2.5$} \\
				\cmidrule{2-5} 
				& $\theta$ & $\hat{\theta}_{IRLS}$ & $\hat{\theta}_{Opt_1}$ &$\hat{\theta}_{Opt_2}$ \\
				& & (sd) & (sd) &(sd) \\
				\midrule
				$\beta_0$  & \textbf{1} &\textbf{1.02}&\textbf{1.02}&\textbf{0.73}  \\
				& &(0.13)&(0.13)&(0.16) \\
				$\beta_1$& \textbf{3} &\textbf{3.09}&\textbf{3.09}&\textbf{2.43}\\
				& & (0.22) & (0.23) &(0.23) \\
				$\beta_2$&\textbf{-3} &\textbf{-3.10}&\textbf{-3.10}&\textbf{-2.42}\\
				& & (0.20) & (0.20) & (0.24) \\ 
				$\beta_3$&\textbf{2} &\textbf{2.17}&\textbf{2.17}&\textbf{1.76}\\
				& & (0.39) & (0.39) & (0.37) \\
				$\beta_4$&\textbf{-2}&\textbf{-2.06}&\textbf{-2.06}&\textbf{-1.61}\\
				& & (0.17) & (0.17) & (0.16) \\
				$log(\nu)$ &\textbf{0.91}&\textbf{0.95}&\textbf{0.94} &\textbf{ 0.69}\\
				& & (0.06) & (0.06) & (0.10) \\
				\bottomrule
			\end{tabular}
			
		\end{minipage}
		\hfill
		\begin{minipage}[b]{0.4\linewidth}
			\small
			\begin{tabular}{ccccc}
				\toprule
				& \multicolumn{4}{c}{$\nu=4$} \\
				\cmidrule{2-5} 
				& $\theta$ & $\hat{\theta}_{IRLS}$ & $\hat{\theta}_{Opt_1}$ &$\hat{\theta}_{Opt_2}$ \\
				& & (sd) & (sd) &(sd) \\
				\midrule
				$\beta_0$  & \textbf{2} &\textbf{2.00}&\textbf{2.01}&\textbf{1.01}  \\
				&  &(0.17)&(0.17)&(0.17)  \\
				$\beta_1$& \textbf{3} &\textbf{3.06}&\textbf{3.07}&\textbf{1.85}\\
				&  &(0.23)&(0.24)&(0.16)\\
				$\beta_2$&\textbf{-3} &\textbf{-3.08}&\textbf{-3.09}&\textbf{-1.88}\\ 
				& &(0.20)&(0.20)&(0.19)\\
				$\beta_3$&\textbf{4} &\textbf{4.28}&\textbf{4.30}&\textbf{2.71}\\
				& &(0.45)&(0.46)&(0.49)\\
				$\beta_4$&\textbf{-4}&\textbf{-4.08}&\textbf{-4.10}&\textbf{-2.38}\\
				&&(0.23)&(0.23)&(0.23)\\
				$log(\nu)$ &\textbf{1.38}&\textbf{1.41}&\textbf{1.40} &\textbf{ 0.86}\\
				&&(0.05)&(0.05) & (0.09)\\
				\bottomrule
			\end{tabular}
		\end{minipage}
		\caption{Comparison of the estimated parameters from three methods ($\hat{\theta}_{IRLS}, \hat{\theta}_{Opt_1}, \hat{\theta}_{Opt_2}$). Results are obtained using 20 bootstrap replications. $\theta$ denotes the true parameter values. Values in parentheses are standard errors.}
		\label{tab:sim-res}
	\end{table}
	%
		% % Figure 2
	\begin{figure}[!htbp]		
		\begin{minipage}{0.9\linewidth}
			\centering
			\includegraphics[height=0.4\textheight,width=0.9\textwidth]{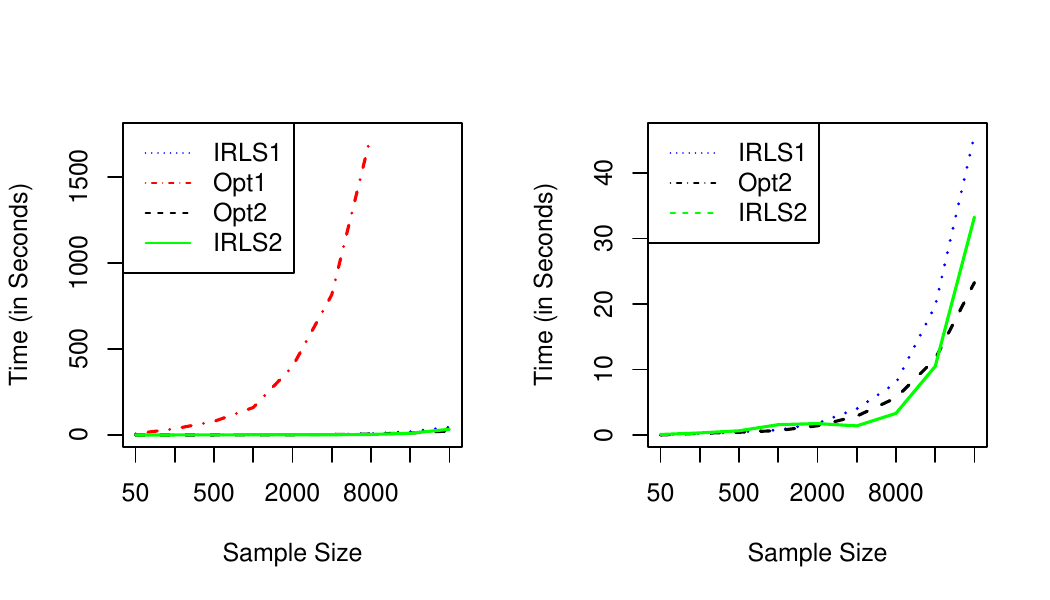}
			\caption{Comparison of the methods in terms of their computational timings for a data with $\nu=0.5$ with increasing sample sizes. While $IRLS_1$ is initialized at $\nu=0.2$, $IRLS_2$ is initialized with a $\nu$ computed from a sample model. Right panel removes $Opt_1$ to provide clearer separation of other methods.}
			\label{fig:odcomp}
		\end{minipage}
		\vfill
		\begin{minipage}{0.9\linewidth}
			\centering
			\includegraphics[height=0.4\textheight,width=0.9\textwidth]{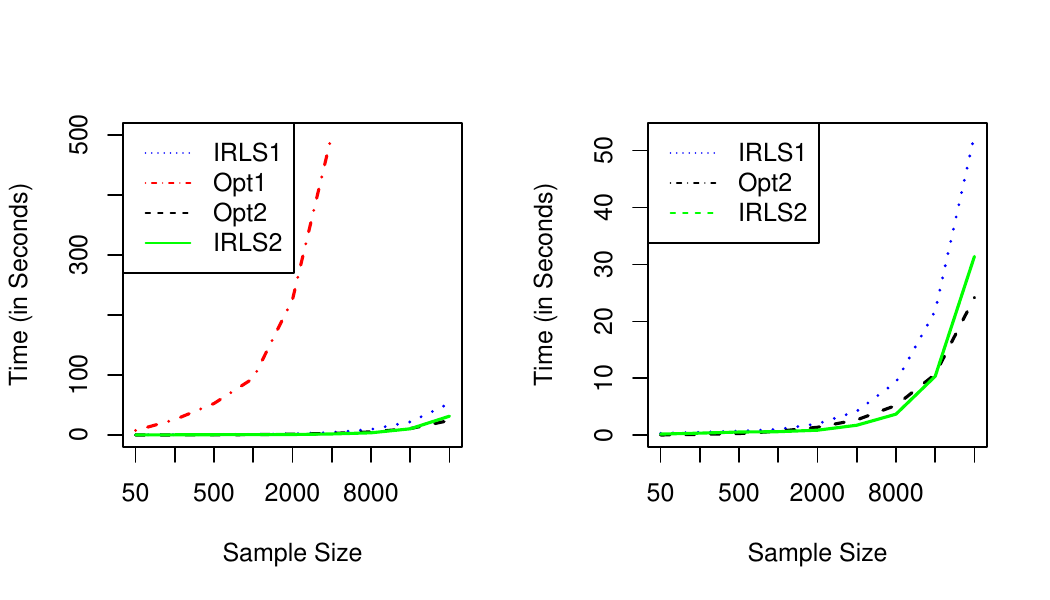}
			\caption{Comparison of the methods in terms of their computational timings for a data with $\nu=4$ with increasing sample sizes. While $IRLS_1$  starts at $\nu=0.2$, $IRLS_2$ starts with a $\nu$ computed from a sample model. Right panel removes $Opt_1$ to provide clearer separation of other methods.}
			\label{fig:udcomp}
		\end{minipage}
	\end{figure}

	For a couple of models from our simulation study $(\nu=0.5, 4)$ we compared the computation times of $IRLS$ with both $Opt_1$ and $Opt_2$ with increasing sample sizes. It is well known that the convergence speed of the IRLS algorithm depends on its starting value. We therefore take a sub sample of data and then run the algorithm to get an approximated value for $\nu$ and feed it as the initial value for the estimation using the full data. We call this $IRLS_2$ and use $IRLS_1$ to denote the original algorithm which always starts at $\nu=0.2$. 
	
	The computation times for the four methods are shown in Figures \ref{fig:odcomp} and \ref{fig:udcomp}. The computation times for $IRLS_2$ include the time for estimating the model for the sub sample to obtain an initial value for the parameter estimates. From the results, it can be observed that $Opt_2$ is  superior to $IRLS_1$, $IRLS_2$ and $Opt_1$.  $Opt_1$ is painfully slow and often takes many minutes where both $IRLS_2$ and $Opt_2$ take a few seconds. As  expected, the $IRLS_2$ algorithm performs much faster than the original $IRLS_1$ and at times it even works faster than $Opt_2$. It is also worth mentioning  that while $Opt_2$ is very fast, it has some issues when there is under dispersion as we have already seen in the simulation results\footnote{Although not reported here, we also examined the performance of the WLS formulation of \citet{sel08} which was described in Section \ref{sec:cmp-reg}. Most of the time, their algorithm converged to the wrong value, typically with $\nu$ close to zero irrespective of the true $\nu$ value used to simulate the data.}.
	
	Ideally, we would want theoretical computation times which can provide a more rigorous comparative study. For the IRLS algorithm, it is easy to obtain the theoretical computation times given the number of iterations needed for the convergence, because for each IRLS iteration we fit two least squares models and we know the theoretical computation times for the least squares regression. However, both  $Opt_1$ and $Opt_2$ algorithms use external optimization functions to optimize the likelihood. The theoretical computation times for these algorithms are not easy to obtain because of the complicated nature of the algorithms and also due to the computations involved in estimating the normalizing constant in the CMP distribution.

	\section{A CMP Generalized Additive Model}\label{sec-GAM}
	\subsection{Background}
	
	A generalized additive model (GAM) \citep{has90} is a
        generalized linear model (GLM; \cite{mcc89}) with a linear
        predictor involving smooth functions of covariates:
	\begin{eqnarray}
          g\{E[y_i]\} &=& \boldsymbol{x}_i^* \boldsymbol{\theta}^* + \sum_{j=1}^{p} f_j(x_{ij}), \quad i=1, \ldots, n
	\end{eqnarray}
	where $g(\cdot)$ is a smooth monotonic and twice
        differentiable link function, $\boldsymbol{x}_i^*$ is the
        $i$th row of $X^*$, which is the model matrix for the
        parametric model components, $\boldsymbol{\theta}^*$ is the parameter  vector, and $f_j$ are the smooth functions
        of the covariate $\boldsymbol{x}_j$ and they are subject to identifiability constraints, such as $\sum_{i=1}^{n}f_j(x_{ij})=0$ for all $j$.  There exist multiple
        methods for estimating the smooth functions $f_j$
        \citep{has90,has86,woo06,rup03,ramsay2006functional,
          gu2013smoothing}. Among those the two most popular approaches
        that use spline bases are smoothing splines \citep{has86,has90} and
        penalized splines \citep{woo06}.

        The smoothing splines approach uses the \emph{backfitting}
        algorithm to estimate the smooth functions. The algorithm can
        be used within the IRLS framework by incorporating another
        inner loop to estimate smooth functions at every
        iteration. The \emph{backfitting} algorithm is elegant as it
        has the flexibility to incorporate a wide variety of smoothing
        methods for component estimation. The convergence of the
        backfitting algorithm and its related properties can be found
        in \cite{buj89}.  However, as suggested by \citet{woo06,gu91},
        it is not easy to efficiently integrate the estimation of the
        smoothing parameter into the model estimation framework. Traditional methods such as cross validation are often
        prohibitive because of the high computational cost involved in
        the search for multiple smoothing parameters.
	
		The penalized splines approach has become a popular choice for
        fitting additive models due to the availability of a variety of
        methods with efficient implementations \citep{woo07}. The idea is to represent each $f_j$ with  intermediate rank spline-type basis expansions, in which case the model becomes the GLM. In order to avoid overfitting, the model is estimated by penalized likelihood maximization. In practice, the penalized maximum likelihood is maximized by penalized iteratively re-weighted least squares (P-IRLS). In particular, the  GAM is fitted by iteratively minimizing 
        \begin{eqnarray}
	        \parallel\sqrt{W^{(k)}}(T^{(k)}-X\boldsymbol{\beta})\parallel^2 + \sum_{j} \eta_j\boldsymbol{\beta}^T S_j \boldsymbol{\beta} \text{ w.r.t. } \boldsymbol{\beta}.
        \end{eqnarray}
         $T^{(k)}$ denotes the adjusted response variable and $W^{(k)}$ denotes the weights at the $k$th iteration of the P-IRLS algorithm. The $S_j$ are matrices of known coefficients such that $\boldsymbol{\beta}^TS_j\boldsymbol{\beta}$ measures the roughness of $f_j$. The $\eta_j$ are smoothing parameters that control the trade-off between fit and smoothness and their selection can be achieved by minimizing the Generalized Cross Validation (GCV) score, AIC, or another criterion \citep{woo06,wood2004stable,wood2008fast}.
         
         There are two types of computational methods available for the estimation of $\eta_j$. (i) \emph{Performance iteration} uses the fact that at each P-IRLS step a working penalized linear model is estimated and the smoothing parameter estimation can be performed on each such working model. (ii) In \emph{outer iteration} the P-IRLS algorithm is iterated to convergence for each trial set of smoothing parameters and the GCV or AIC scores are only evaluated on convergence \citep{woo06}.  
        
        \subsection{Implementation of the CMP Generalized Additive Model}
	Similar to the CMP regression, the CMP generalized additive model can be formulated as
	\begin{eqnarray}
		\ln(\lambda_i) &=& \boldsymbol{x}_i^* \boldsymbol{\theta}^* + \sum_{j=1}^{p} f_j(x_{ij}), \\
		\ln(\nu_i) &=& \boldsymbol{z}_i^* \boldsymbol{\delta}^* + \sum_{j=1}^{k} m_j(z_{ij}),
		\label{eqn: cmp-gam}
	\end{eqnarray}
	for $i=1, \ldots,n$, where $\boldsymbol{\theta}^*$ and $\boldsymbol{\delta}^*$ are the parameter vectors for the parametric part of $\ln(\lambda_i)$ and $\ln(\nu_i)$ respectively. The smooth functions $f_j$ and $m_j$ are the smooth functions for the covariates $\boldsymbol{x}_j$ and $\boldsymbol{z}_j$ and are subject to the identifiability constraints. For the sake of easy presentation, we omit the strictly parametric part from now onwards.
	
	We consider the \emph{performance iteration} method  as it is very efficient and computes faster than the \emph{outer iteration} method.  Although not common, there is some evidence that \emph{performance iteration} faces some convergence issues because the objective function for the smoothing parameters keeps changing with every iteration of P-IRLS \citep{woo06,woo11}. In contrast, the \emph{outer iteration} method is more robust to convergence related issues but usually takes longer time to compute. More importantly, on convergence, it requires some derivatives to estimate the smooth parameters. The only way to get derivatives for the CMP distribution is to use numerical derivatives and they are often prone to errors due to the normalizing constant in the likelihood which is an infinite series. 
	For this reason, we consider only the \emph{performance iteration} method to estimate CMP GAM and leave the  \emph{outer iteration} method for future research.      
	
	Now, each P-IRLS iteration involves minimizing the  following two objective functions:
	\begin{eqnarray}
		  \parallel\sqrt{W_1^{(k)}}(T_1^{(k)}-X\boldsymbol{\beta})\parallel^2 + \boldsymbol{\beta}^TH_1\boldsymbol{\beta}  +\sum_{j} \eta_{1j}\boldsymbol{\beta}^T S_{1j} \boldsymbol{\beta} \text{ w.r.t. } \boldsymbol{\beta}, \\
		    \parallel\sqrt{W_2^{(k)}}(T_2^{(k)}-Z\boldsymbol{\gamma})\parallel^2 + \boldsymbol{\gamma}^TH_2\boldsymbol{\gamma} + \sum_{j} \eta_{2j}\boldsymbol{\gamma}^T S_{2j} \boldsymbol{\gamma} \text{ w.r.t. } \boldsymbol{\gamma}.
		    \label{eqn: pwls}
	\end{eqnarray} 
	 $W_1$, $W_2$ are the weight matrices, $\eta_{1j}$, $\eta_{2j}$ are the smooth parameters  for the $\ln(\lambda_i)$ and $\ln(\nu_i)$ models. The matrices $H_1$, $H_2$ are fixed positive semi definite penalty matrices which allow for multiple extensions to the GAMs such as ridge penalties under suitable constraints \citep{wood2004stable}. $T_1^{(k)}$ and $T_2^{(k)}$ are defined similarly as in the Algorithm 1.
	 
	 Given the smoothing parameters $\eta_{1j}, \eta_{2j}$ the objective functions in (\ref{eqn: pwls}) are solved using any penalized least squares type of methodology. However, the smoothing parameters need to be estimated here. We consider the GCV method, in which the smoothing parameters are chosen to minimize 
	\begin{eqnarray*}
		V_{1g} = \frac{n\parallel T_1-A_1T_1 \parallel^2}{[\text{tr}(I-\gamma_1
		A_1)]^2}, \quad
	V_{2g} = \frac{n\parallel T_2-A_2T_2 \parallel^2}{[\text{tr}(I-\gamma_2
		A_2)]^2},
	\end{eqnarray*}
    respectively. $A_1, A_2$ are the influence or hat matrices of the corresponding fitting problems and they depend on the smoothing parameters. The parameters $\gamma_1, \gamma_2$ are sometimes used to inflate the GCV objective function to make sure that the models are smoother \citep{wood2004stable,cha98}. There are efficient algorithms available in the \emph{mgcv}  \citep{woo07,wood2008fast,wood2004stable} package (e.g. \emph{magic} function) to estimate the smoothing parameters along with the model parameters and we use them in our implementation for CMP GAM.
    
    The inference for the spline regression coefficients in GAM is developed using a bayesian view of the smoothing process, in which the smoothing penalties are induced by  improper Gaussian priors on $\boldsymbol{\beta}$, $\boldsymbol{\gamma}$ and $\hat{\boldsymbol{\beta}}$, $\hat{\boldsymbol{\gamma}}$ are also the modes of the posterior densities of $\boldsymbol{\beta}$, $\boldsymbol{\gamma}$   \citep{wahba1983bayesian,silverman1985some,nychka1988bayesian}. Please refer to \citet{wood2006confidence,wood2012p} for more details. Based on the results from \citet{wood2006confidence,woo11}, the large sample posterior distribution for the regression spline coefficients in CMP GAM is
    \begin{eqnarray*}
    \boldsymbol{\beta}\vert \boldsymbol{\eta}_1,Y \sim \mathcal{N}(\hat{\boldsymbol{\beta}},\Sigma_\beta), \quad
    \boldsymbol{\gamma}\vert \boldsymbol{\eta}_2,Y \sim \mathcal{N}(\hat{\boldsymbol{\gamma}},\Sigma_\gamma), 
    \end{eqnarray*}  
    where 
    $\Sigma_{\beta}=(X^TW_1X+\sum_{j=1}^{p}\eta_{1j}S_{1j})^{-1}\phi_1$, $\Sigma_{\gamma}=(Z^TW_2Z+\sum_{j=1}^{q}\eta_{2j}S_{2j})^{-1}\phi_2$ with $\phi_1$, $\phi_2$ as the estimated scale parameters and $W_1$, $W_2$ are the weight matrices at the convergence of the P-IRLS algorithm. 
    
    From a practitioner point of view, sometimes it is required to check whether a particular smooth function is significant in the model or can be discarded. More formally, to test the null hypotheses that $f_j=0$ or $m_j=0$ for any j, \citet{wood2012p} proposed the  Wald test statistic ($wt_r$), which under the null hypothesis and with a large sample follows a chi-square distribution.  The Wald test statistic is defined as $wt_{r}=\hat{f}_j^TV_{f_j}^{r-}\hat{f}_j$ where $V_{f_j}^{r-}$ is the rank-$r$ pseudo-inverse of $V_{f_j}=\boldsymbol{x}_jV_{\beta}\boldsymbol{x}_j^T$. The authors have suggested that naive choices of $r$ lead to low power or an incorrect null distribution for $p$-values and using the effective degrees of freedom for $r$ is a better choice. For more details please refer to \citet{wood2012p}.   
    
    The inference procedure proposed for CMP GAM is based on the marginal likelihood. If needed, one could also develop the conditional inference framework by suitably adjusting the covariance matrices. However, it is not guaranteed that the rank-r pseudo inverse remains optimal. Further, it is also possible that after the correction, the covariance matrix may not be positive definite. For this reason, we only use the inference procedure based on marginal likelihood. In practice, we found that although there are some changes in p-values, the  results remain same whether we use inference procedure developed based on marginal likelihood or conditional likelihood.

	\section{Simulation Study for the CMP Generalized Additive Model}
	\label{sec-simulation-gam}
	
	We conducted a simulation study to evaluate the usefulness of the CMP GAM for fitting non linear terms. We consider two examples for our simulation study; in the first example we choose one fixed value for $\boldsymbol{\nu}$ such as $0.5$ or $2.5$ and in the second example   we simulate $\boldsymbol{\nu}$ using a nonparametric smooth function. While the first example is considered to compare CMP GAM with other models such as NB GAM, Poisson GAM and CMP GLM (CMP Regression), the second example is considered to showcase the flexibility of CMP GAM allowing dispersion to vary non linearly across observations.
	
	\subsection{Example 1} \label{sub-sec-simgam-1}
	Inspired by the four uni-variate example  \cite{gu91,woo07}, we simulated data from a CMP GAM, with sample size 500, as following:
	\begin{itemize}
		\item Simulate $x_1,x_2,x_3$ and $x_4$ from a standard uniform $U[0,1]$ distribution.
		\item Consider the functions $f_1=\sin(\pi x_1), f_2=\exp(x_2), f_3=0.02x_3^2(1-x_3)+(0.5x_3)^2 (1-x_3)^3$ and $f_4=x_4$.
		\item  Calculate $f=af_1(x_1)+bf_2(x_2)+cf_3(x_3)$, where $a,b,c$ are pre-specified constants.
		\item Set $\lambda= \exp(f)$ and simulate data for a fixed $\nu$.
	\end{itemize}
	
	Although we simulated 4 covariates, we have used only the first three $(x_1,x_2,x_3)$ to compute  $\lambda$, which is used to simulate the dependent variable. We still use $x_4$ to estimate the models. This allows us  to check how the proposed method deals with over-specification. In an ideal case, the estimated model would identify $x_4$ as non significant. We consider two different values for $\nu$ $(=0.5,2)$ to capture both over dispersion and under dispersion scenarios. Unlike standard GLMs, simulating data from the CMP distribution requires a careful consideration of parameters $\lambda, \nu$. The scale of the dependent variable is in the range of $\lambda^{1/\nu}$. If we choose $\nu<1$ (over dispersion), then we must consider smaller values for $\lambda$ otherwise we will generate only very large counts (e.g. 1500, 2000), resulting in a data set that is less useful for illustrating count data models. Similarly, for  $\nu>1$, we must consider larger values for $\lambda$ to avoid generating only very small counts such as 0,1,2, resulting in an extremely under dispersed data set that is less useful for comparing against count models such as Poisson or Negative Binomial GAM. To avoid the aforementioned problems, we choose two sets of different values for constants $a,b,c$ such as $\{0.2, 0.5, -0.5\}$ for $\nu=0.5$ and $ \{1, 1, 1\}$ for $\nu=2.5$.
	
	We also considered a Poisson GAM and a CMP GLM model for comparison with the CMP GAM. We used the \emph{mgcv}  \citep{woo07} package in R to estimate the Poisson GAM and NB GAM as they are also implemented using penalized splines. Although the current implementations of both Poisson GAM and NB GAM do not use the P-IRLS algorithm by default, we specifically used P-IRLS (\emph{performance  iteration}) algorithm to estimate these models to provide a fair comparison. 
	
	For each of the two scenarios ($\nu=0.5,2.5$) we used 50 bootstrap replications and recorded the significance levels for each nonparametric term in the model. We also recorded and compared their $AIC$. While the model equation for an additive model is $y \sim s(x_1)+s(x_2)+s(x_3)+s(x_4)$, where $s(\cdot)$ is the smooth function,  the model for the CMP GLM is  $y \sim x_1+x_2+x_3+x_4 $ without any smooth functions for covariates. 
	
	The simulation results are summarized in Table \ref{tab:gamnu}.  The top table (a) describes the results for $\nu=0.5$ and the bottom table (b) describes the results for $\nu=2.5$. Since we estimated 50 models for each bootstrap data, we plotted the AIC values for each data for all the models in Figure \ref{fig:aic-gam} in the Appendix. From the top plot, for $\nu=0.5$, it can be seen that   $AIC$ for CMP GAM is consistently better than the $AIC$ for both  Poisson GAM and CMP GLM. The $AIC$ values for  NB GAM are closer to the $AIC$ values for CMP GAM which indicates that their fits are reasonably close. While both CMP and NB GAMs declared $s(x_3)$ as non significant, Poisson GAM identified it as significant at least half the times. Similarly, for the smooth term $s(x_4)$, which is not part of the true model, Poisson GAM declared it as significant at least half the times. Ideally, CMP GLM should not produce any significant coefficients because of the nonlinear terms in the true model. However, it can be observed that $x_2$ is significant and this is because the function $f_2=\exp(x_2)$ is approximately equal to %\approx 
    $1+x_2$ (because $x_2 \in [0,1]$).

		%%%%
	\begin{table}%[!htbp]
		\begin{minipage}{.9\linewidth}
			\centering
			\begin{tabular}{ccccc}
				\toprule
				& \multicolumn{4}{c}{\#Significant (of 50 bootstraps) } \\
				& \multicolumn{4}{c}{ $(\le 0.001,\le 0.01, \le 0.5, \le 0.1, n.s)$} \\
			%	\cmidrule{2-4} 
				& $cmp-gam$ & $poisson-gam$ & $nb-gam$& $cmp-glm$  \\
			%	\midrule
				%$\beta_0$  & \textbf{2} &\textbf{2.00}&\textbf{2.01}&\textbf{1.01}  \\
				%&  &(0.17)&(0.17)&(0.17)  \\
				$s(x_1)$& (\textbf{46},4,0,0,0) &(\textbf{50},0,0,0,0)& (\textbf{42},8,0,0,0) &(0,0,0,1,\textbf{49})\\
				$s(x_2)$&(\textbf{50},0,0,0,0) & (\textbf{50},0,0,0,0)&(\textbf{50},0,0,0,0)&(\textbf{50},0,0,0,0)\\ 
				$s(x_3)$&(0,1,1,4,\textbf{44}) &(1,4,13,6,\textbf{26})&(0,0,3,3,\textbf{44})&(0,0,2,3,\textbf{45})\\
				$s(x_4)$&(0,0,1,5,\textbf{44}) &(2,8,8,7,\textbf{25})&(0,0,1,3,\textbf{46}) &(0,1,3,2,\textbf{44})\\
				\hline
				& \multicolumn{4}{c}{Estimation and Fit} \\
			%	\cmidrule{2-4}
				& $cmp-gam$ & $poisson-gam$ & $nb-gam$& $cmp-glm$  \\
			%	\midrule
				$\nu$ or $\theta$ &\textbf{-0.78}(0.18)& & \textbf{9.92}(2.05) &\textbf{-0.81}(0.08)\\
				$AIC$ &\textbf{2726.53}&\textbf{2830.73}& \textbf{2734.18}&\textbf{2757.16} \\
			%	\bottomrule
				
			\end{tabular}
			\label{tab:gamnuhalf}
			\caption*{(a) For $\nu=0.5$.}		
		\end{minipage}
		\vfill
		\begin{minipage}{0.9\linewidth}
			\centering
			\begin{tabular}{ccccc}
				\toprule
				& \multicolumn{4}{c}{\#Significant (of 50 boostraps) } \\
				& \multicolumn{4}{c}{ $(\le 0.001,\le 0.01, \le 0.5, \le 0.1, n.s)$} \\
			%	\cmidrule{2-4} 
				& $cmp-gam$ & $poisson-gam$ & $nb-gam$& $cmp-glm$  \\
			%	\midrule
				%$\beta_0$  & \textbf{2} &\textbf{2.00}&\textbf{2.01}&\textbf{1.01}  \\
				%&  &(0.17)&(0.17)&(0.17)  \\
				$s(x_1)$& (\textbf{50},0,0,0,0) &(\textbf{33},15,2,0,0)& (\textbf{50},0,0,0,0) &(0,0,0,2,\textbf{48})\\
				$s(x_2)$&(\textbf{50},0,0,0,0) & (\textbf{50},0,0,0,0)&(\textbf{50},0,0,0,0) &(\textbf{50},0,0,0,0)\\ 
				$s(x_3)$&(0,0,2,5,\textbf{43}) &(0,0,0,0,\textbf{50})&(0,0,2,5,\textbf{43}) &(0,0,2,2,\textbf{46})\\
				$s(x_4)$&(0,0,0,8,\textbf{42}) &(0,0,0,0,\textbf{50}) & (0,0,1,8,\textbf{41})  &(0,0,2,2,\textbf{46})\\
				\hline
				& \multicolumn{4}{c}{Estimation and Fit} \\
			%	\cmidrule{2-4}
				& $cmp-gam$ & $poisson-gam$ & $nb-gam$ & $cmp-glm$  \\
			%	\midrule
				$\nu$ or $\theta$&\textbf{0.84}(0.09)& &\textbf{10000}(0)  &\textbf{0.77}(0.05)\\
				$AIC$ &\textbf{1438.74}&\textbf{1565.56}& \textbf{1568.13}&\textbf{1484.84} \\
		%		\bottomrule
			\end{tabular}
			\label{tab:gamnutwohalf}
			\caption*{(b) For $\nu=2.5$.}
		\end{minipage}
		\caption{Comparison of coefficient significance level and fit among CMP GAM, Poisson GAM, NB GAM and CMP GLM with 50 bootstrap replications. The model for gams is $y \sim s(x_1)+s(x_2)+s(x_3)+s(x_4)$ and for the regression $y \sim x_1+x_2+x_3+x_4$. For the Estimation and Fit results the numbers in parenthesis are standard errors. \\ }
		\label{tab:gamnu}
	\end{table}
	
	The results from  Table \ref{tab:gamnu} (b) can be interpreted similarly. Not surprisingly, in terms of the model fit, both Poisson GAM and NB GAM are very close but not better than the CMP GAM which is evident from the bottom plot in Figure \ref{fig:aic-gam} in the Appendix. This is because of their inability to handle under dispersion.    

	\subsection{Example 2} \label{sub-sec-simgam-2}
	We simulate data similar to the procedure in Section \ref{sub-sec-simgam-1}. The procedure is as follows:
	\begin{itemize}
		\item We consider the same functions  for $f_1$, $f_2$ and $f_3$ but a different function for $f_4=2x_4-x_4^2$.
		\item Calculate $f=f_2+f_3+2f_4$ and set $\boldsymbol{\lambda}=\text{exp}(f)$.
		\item Calculate $g=f_1$ and set $\boldsymbol{\nu}=\text{exp}(g)$.
		\item Simulate $y_i$ from $CMP(\lambda_i,\nu_i)$.
	\end{itemize}
Since the dispersion parameter is generated using smooth function, the observations will have different dispersions. In the simulated data, the dispersion $(\nu_i)$ varies from 0.7 to 2.5. We now use the data and fit two different models. We estimate $\boldsymbol{\nu}$ non parametrically in the first model and parametrically in the second model. 

Table \ref{tab:gamnusmooth} contains the significance levels results from the above mentioned two models using 50 bootstrap replications. As expected modeling $\boldsymbol{\nu}$ non parametrically yields better fit.  As seen from the results in Section \ref{sub-sec-simgam-1}, the parametric model is not able to capture the underlying nonlinear function for $\boldsymbol{\nu}$.  
	
	%%%
	\begin{table}%[!htbp]
		\centering
			\begin{tabular}{ccc}
				\toprule
				& \multicolumn{2}{c}{\#Significant (of 50 bootstraps) } \\
				& \multicolumn{2}{c}{ $(\le 0.001,\le 0.01, \le 0.5, \le 0.1, n.s)$} \\
				\cmidrule{2-3} 
				& \multicolumn{2}{c}{$cmp-gam$}  \\
			 \cmidrule{2-3} 
				& $m_1$ & $m_2$  \\
				\midrule
				$\boldsymbol{\lambda}$ model: & &\\
				$s(x_4)$& (\textbf{50},0,0,0,0) &(\textbf{50},0,0,0,0) \\
				$s(x_2)$&(\textbf{50},0,0,0,0) & (\textbf{50},0,0,0,0)\\ 
				$s(x_3)$&(0,1,3,1,\textbf{45}) &(1,4,9,5,\textbf{31})\\
				\\
				$\boldsymbol{\nu}$ model: & & \\
				
				$\gamma_0$& (\textbf{50},0,0,0,0) &(\textbf{50},0,0,0,0)  \\
				$s(x_1)$ or $\gamma_1$& (\textbf{50},0,0,0,0) &(2,1,8,2,\textbf{37}) \\
				
				\bottomrule
				
			$AIC$ &\textbf{1968.62}&\textbf{2891.60}\\
		\bottomrule
			\end{tabular}
		\caption{Comparison of coefficient significance levels and fit between CMP GAMs with two different models for $\boldsymbol{\nu}$ based on 50 bootstrap replications. The models for $\boldsymbol{\nu}$ are $m_1=\sim s(x_1); m_2=\sim x_1$. The model for $\boldsymbol{\lambda}$ is $\sim s(x_4)+s(x_2)+s(x_3) $ for both $m_1$ and $m_2$.}
			\label{tab:gamnusmooth}
		\end{table}

	\section{Example: Modeling Data from Bike Sharing Systems}\label{sec-bikeshar}
To illustrate the use of the CMP GAM on real data, we model data from a bike sharing application. Bike sharing systems are a new generation of traditional bike rental services where the entire process that includes membership, bike rental and return has become automated. Through these systems, users are able to easily rent a bike from a particular station and return it to another location. There is a need for bike sharing programs to effectively understand the factors that influence demand so that they can better maintain inventory, schedule repairs, and manage resources. We therefore use a GAM model in this context. 

Data collected by bike sharing systems typically include information on each trip taken (time stamps and locations of rental and return) and sometimes also information on the rider. Several datasets from real bike sharing systems are publicly available. We use the data available from \citet{fan20} on rides in Washington, DC between 2011-2012. The data is available in two formats: daily and hourly number of rentals. We chose the hourly data as it is more complex and better illustrates the new models that we introduce. The data includes information about the number of rides by casual users and registered users for every hour during the years 2011 and 2012. In addition, it also includes external information such whether the hour is on a weekday, a working day or a holiday, the weather situation (clear/cloudy/rainy), temperature, and wind speed. These external factors are considered detrimental to the demand for bikes.  For a full list of attributes and their descriptions please refer to the Table \ref{tab:bikeshar-attr} in the Appendix.
    
	 We considered the following model for both counts of casual and registered users:
	\begin{align*}
		\ln(\boldsymbol{\lambda}) &= \beta_0+\beta_1 hour+\beta_2 holiday+ \beta_3 weekday+\beta_4 weathersit+ \\
		&s(atemp)+s(hum)+s(windspeed)+s(day) \\
		\ln(\boldsymbol{\nu}) &=\gamma_0.
	\end{align*} 
	Since the attributes \emph{atemp} and \emph{temp} are highly correlated, we included only \emph{atemp} in the model. We kept control variables such as \emph{holiday} and \emph{weathersit} as parametric terms and included other continuous variables of interest such as \emph{hum} and \emph{windspeed}  as nonparametric components.

	For this study we have only considered the January 2012 data. The same analysis can be repeated for every month or for every season. Since we have two dependent variables of interest we fit two models to this data. The first model is for the number of hourly rentals for registered users and the second model is for the number of hourly rentals for casual users. For the sake of comparison we  also fit a Poisson GAM  and NB GAM using the  \emph{mgcv} \cite{woo07} package. As in the simulation study in Section \ref{sec-simulation-gam}, we made sure that both the Poisson and NB GAMs are estimated via the \emph{performance iteration} algorithm. 
    
    The coefficient significance results are described in Table \ref{tab:bike-jan-reg-cas}. For brevity we did not include the coefficient significance results for the parametric components.  We only reported the results for the non parametric components and the findings from these are similar to the findings from the parametric part. From the registered users results in Table \ref{tab:bike-jan-reg-cas}, it can be observed that while in CMP GAM all the smooth variables are significant, in CMP and NB GAM the smooth variable \emph{windspeed} is not significant. Since the data exhibits high levels of dispersion, which is evident from $\gamma_0$ ($<0$ over dispersion; $>0$ under dispersion), the significance results from NB GAM and CMP are similar.  Based on the $AIC$ values, not surprisingly, both CMP and NB GAM fit better than Poisson GAM.

	\begin{table}[!htbp]
%		\centering
		\begin{minipage}{0.4\linewidth}
		%	\begin{center}
				\begin{tabular}{l c c c }
					\hline
					& \multicolumn{3}{c}{Registered Users} \\
					& $cmp-gam$ & $poisson-gam$ & $nb-gam$ \\
					& (edf) & (edf) & (edf) \\
					\hline
					%(Intercept)    &  & & \\
					s(day) & $5.63^{**}$ & $7.91^{***}$ & $3.40^{**}$\\
					s(atemp)& $6.85^{***}$ & $8.86^{***}$ & $7.11^{***}$   \\
					s(hum)  & $7.96^{**}$ & $8.97^{***}$ & $8.31^{***}$\\
					s(windspeed) & $2.31$ &	 $8.90^{***}$		& $1.00$\\	
					\hline 
					$\gamma_0$ or $\theta$  & $-3.03^{***}$ & & $3.82^{***}$ \\ 
					\hline
					AIC    & 7413.55      & 18639.87 & 7563.97     \\
					RMSE  & 49.41 & 49.46 & 59.10 \\
					\hline
					\multicolumn{2}{l}{\scriptsize{$^{***}p<0.001$, $^{**}p<0.01$, $^*p<0.05$}}
				\end{tabular}
				%\caption{Comparison of CMP GAM and Poisson GAM in terms of coefficient significance and fit for \# rentals for registered users in January 2012.}
				%\label{tab:bike-jan-reg}
		%	\end{center}
		\end{minipage}
		\vfill
		\begin{minipage}{0.4\linewidth}
		%	\begin{center}
				\begin{tabular}{lccc}
					\hline
					& \multicolumn{3}{c}{Casual Users} \\
					& $cmp-gam$ & $poisson-gam$ & $nb-gam$ \\
					  & (edf) & (edf)& (edf) \\
					\hline
					% (Intercept)    & & &  \\
				s(day) & $8.90^{***}$ & $8.96^{***}$ & $8.82^{***}$\\
					 s(atemp)	& $1.00^{***}$ & $5.12^{***}$ & $1.00^{***}$   \\
				 s(hum)  & $8.05^{**}$ & $8.81^{***}$& $1.78^{***}$\\
				 s(windspeed) & $1.68$ &	 $6.11^{***}$ & $1.00$	\\		
					\hline 
				$\gamma_0$ or $\theta$  & $-1.36^{***}$ & &$3.20^{***}$ \\ 
					\hline
				 AIC            & 3990.84      & 4713.76 & 4044.07     \\
				 RMSE  & 6.59  & 6.91 & 9.59 \\
					\hline					\multicolumn{2}{l}{\scriptsize{$^{***}p<0.001$, $^{**}p<0.01$, $^*p<0.05$}}
				\end{tabular}
				%\caption{Comparison of CMP GAM and Poisson GAM in terms of coefficient significance and fit, for \# rentals for casual users in January 2012.}
			%\label{tab:bike-jan-cas}
		%	\end{center}	
		\end{minipage}
		\caption{Comparison of CMP GAM, Poisson GAM and NB GAM in terms of coefficient significance and fit for \# rentals for both registered (top) and casual (bottom) users in January 2012.  }
		\label{tab:bike-jan-reg-cas}
	\end{table}
		\begin{figure}[!htbp]
		\begin{minipage}{0.9\linewidth}
			\begin{center}
				\includegraphics[height=0.3\textheight,width=\textwidth]{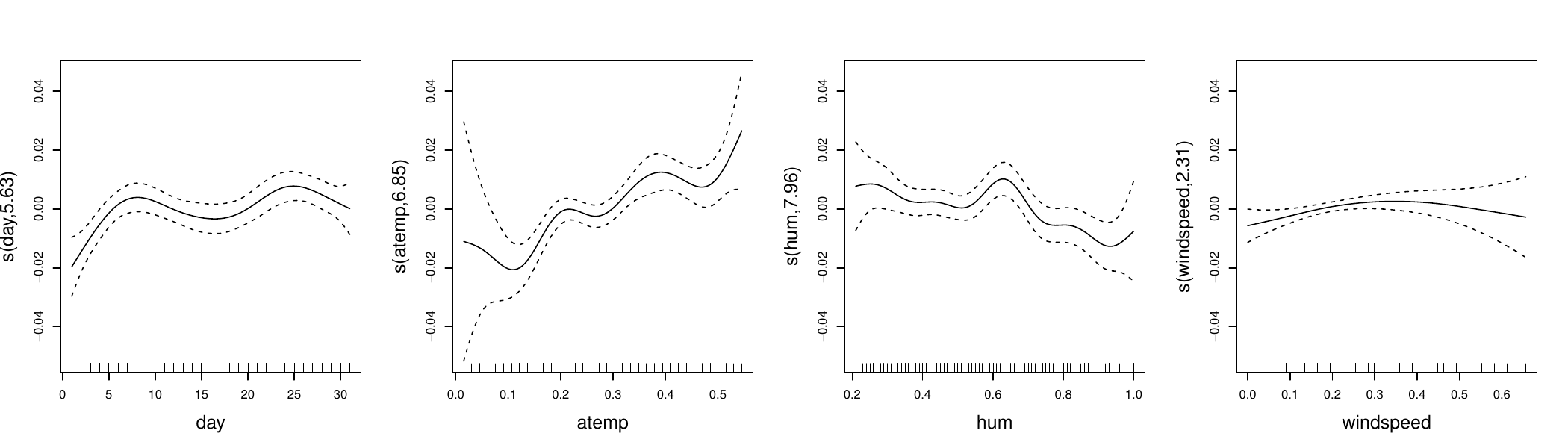}	
				\caption{Partial plots for the smooth variables for CMP GAM. The dependent variable is \# rentals for registered users in January 2012.}
				\label{fig:reg-jan}
			\end{center}
		\end{minipage}
		\vfill
		\begin{minipage}{0.9\linewidth}
			\begin{center}
				\includegraphics[height=0.3\textheight,width=\textwidth]{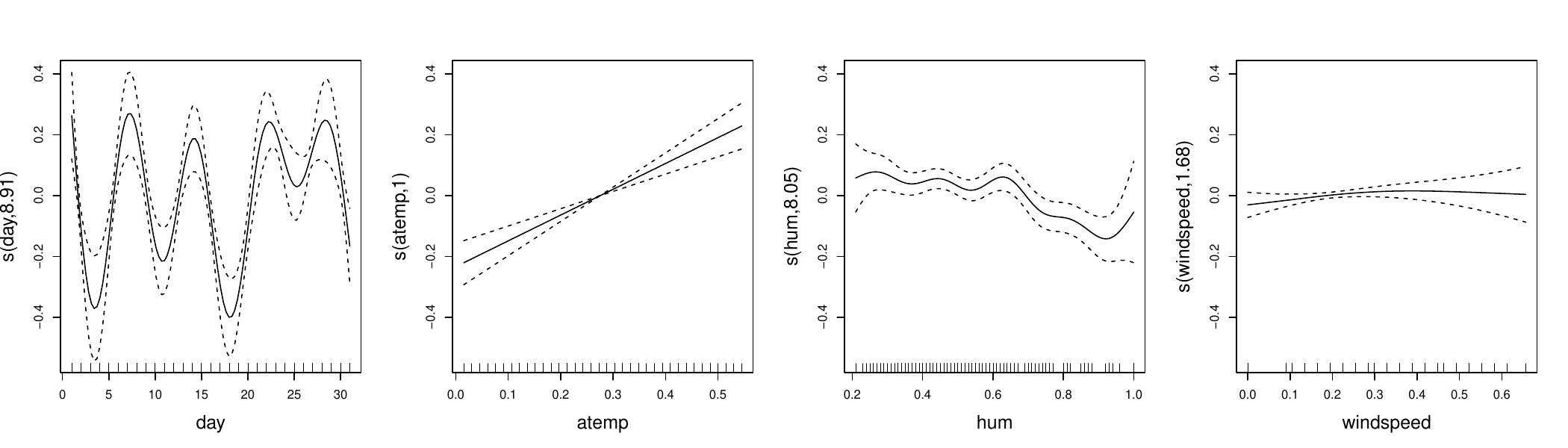}
				
				\caption{Partial plots for the smooth variables for CMP GAM. The dependent variable is \#rentals for casual users in January 2012.}
				\label{fig:cas-jan}
			\end{center}
		\end{minipage}
	\end{figure}
	Similar results are seen in Table \ref{tab:bike-jan-reg-cas} for the  casual users. In Poisson GAM all the smooth variables are significant whereas in CMP GAM  and NB GAM the smooth variable \emph{windspeed} is not significant. Similar to the registered users, the data is over dispersed. These results therefore also highlight potential inference errors when fitting a Poisson GAM to data with excessive dispersion.
	
	To provide more meaningful interpretations we use partial plots. The partial plots for the CMP GAM for the registered users are shown in Figure \ref{fig:reg-jan}. %Similar to the significance table, we have included the plots from Poisson additive model. 
     The smooth variables \emph{day} and \emph{atemp} exhibit an increasing trend while \emph{humidity} exhibits a decreasing trend. This is expected because when there is high humidity people may not show much interest in riding bikes. Further, in a winter month like January, high temperature (or sunny day) encourages people to ride bikes.  The insignificance of \emph{windspeed} is evident from the plot as the smooth curve is close to zero throughout the  range.  To draw more meaningful interpretations we would need more domain knowledge.   

	For casual users one can draw interpretations from the partial plots in Figure \ref{fig:cas-jan}. The results are similar to  Figure \ref{fig:reg-jan} except for  the smooth variable \emph{day}, which shows a cyclical pattern that might indicate the high demand for  bikes  during weekends.

Finally, from an actual model fit perspective, both CMP and NB GAMs perform reasonably well.  Figure \ref{fig:bikeshar-jan-pred} in the Appendix compares fitted values and residuals from CMP, Poisson and NB GAMs.  Since the data is hourly, we plotted the fitted values for every hour. We joined the data points to provide better visualization. We see that for both registered and casual users, the CMP and Poisson GAMs fitted values are close to the actual values while the NB GAM fitted values are not. The identical performance of the CMP and Poisson GAMs in terms of fitted values is expected as they differ only in terms of standard errors rather than point predictions.

In summary, CMP GAM can be a valuable addition for modeling count data. Despite its computational complexity,  CMP GAM is very flexible as it can handle both over dispersion and under dispersion which existing methods fail to handle. Although the bike sharing data did not exhibit under dispersion, there are plenty of data sets that do. Moreover, when the researcher does not know the dispersion type (over or under) prior to modeling, CMP GAM is a safe option.  
	\section{Conclusions and Future Directions}\label{sec-conclude}
	We introduced a flexible estimation framework for estimating a CMP GLM model that is based on the IRLS approach. This framework allows the CMP distribution to join other existing GLMs where IRLS is used for efficient estimation as well as for various modeling enhancements. This framework can be further developed to extend methods such as the lasso.  
	
	While the IRLS algorithm for CMP GLM is computationally intensive compared to an ordinary Poisson model,  the computation time can be reduced by suitably parallelizing some of the computations such as the calculation of  cumulants. Such parallel computing will be beneficial especially with large samples.
	
	In this work we explored fitting additive models with penalized splines. We considered the \emph{performance iteration} method to fit the model as it is based on the P-IRLS algorithm. One possible extension is to develop the \emph{outer iteration} method using  the Newton algorithm. The numerical derivatives required for the Newton algorithm are computationally slower and not  very stable, thereby,requiring new, efficient implementation. 

\section*{Acknowledgements}
The authors were partially funded by research grant 105-2410-H-007-034-MY3 by the Ministry of Science and Technology in Taiwan. The authors would like to thank the AE and the two referees for their numerous suggestions that substantially improved the content and the presentation of the paper. The authors also gratefully acknowledge Prof. Li-Shan Huang and Prof. Satish Iyengar for their valuable feedback and suggestions on earlier versions of this manuscript.

%\section*{\refname}
\bibliographystyle{elsarticle-harv}
	\bibliography{example}

	\newpage
\appendix
\section*{Appendix}
\label{sec-suppl}
\subsection*{R Package}
	\begin{description}		
		\item We created an R-package (\emph{cmp}) with all the methods developed in the paper. The package is available on github and can be installed by running the following R code: 
        \begin{center}
      \texttt{require(devtools)}\\
       \texttt{install\_github("SuneelChatla/cmp")} \\
      \texttt{require(cmp)}
 \end{center}		
	\end{description}
	
\subsection*{The full list of attributes and their descriptions for the Bikesharing data}

\setcounter{table}{0} \renewcommand{\thetable}{A.\arabic{table}}
	\begin{table}[!htbp]
		\centering
		\begin{tabular}{ll}
			\toprule
			Name & Description \\
			\midrule
			\emph{dteday} & date
\\
			\emph{season} & season (1:spring, 2:summer, 3:fall, 4:winter)
\\
			\emph{yr} & year (0: 2011, 1:2012)
\\
			\emph{mnth} & month ( 1 to 12)
\\
			\emph{hr} & hour (0 to 23)
\\
			\emph{holiday} & weather the day is holiday or not (extracted from\\ &\url{http://dchr.dc.gov/page/holiday-schedule}) \\
			\emph{weekday} & day of the week
\\
			\emph{workingday} & if day is neither weekend nor holiday is 1, otherwise is 0.
\\
			\emph{weathersit} & 1= Clear, Few clouds, Partly cloudy \\
				& 2= Mist + Cloudy, Mist + Broken clouds, Mist + Few clouds, Mist\\
				& 3= Light Snow, Light Rain + Thunderstorm + Scattered clouds \\
				& 4= Heavy Rain + Ice Pallets + Thunderstorm + Mist, Snow + Fog \\	
			\emph{temp} & Normalized temperature in Celsius. The values are divided to 41 (max)
\\
			\emph{atemp} & Normalized feeling temperature in Celsius. The values are divided to 50 (max)\\
			\emph{hum}& Normalized humidity. The values are divided to 100 (max)
\\
			\emph{windspeed} &Normalized wind speed. The values are divided to 67 (max)
\\
			\emph{casual} & count of casual users
\\
			\emph{registered} & count of registered users
\\
			\emph{cnt}& count of total rental bikes including both casual and registered \\
			\bottomrule
		\end{tabular}
	\caption{Full list of attributes and their description for the bike sharing data}
	\label{tab:bikeshar-attr}
	\end{table}

	\setcounter{figure}{0} \renewcommand{\thefigure}{A.\arabic{figure}}
	\begin{figure}
	    \centering
	    \includegraphics[height=0.5\textheight,width=1.1\textwidth]{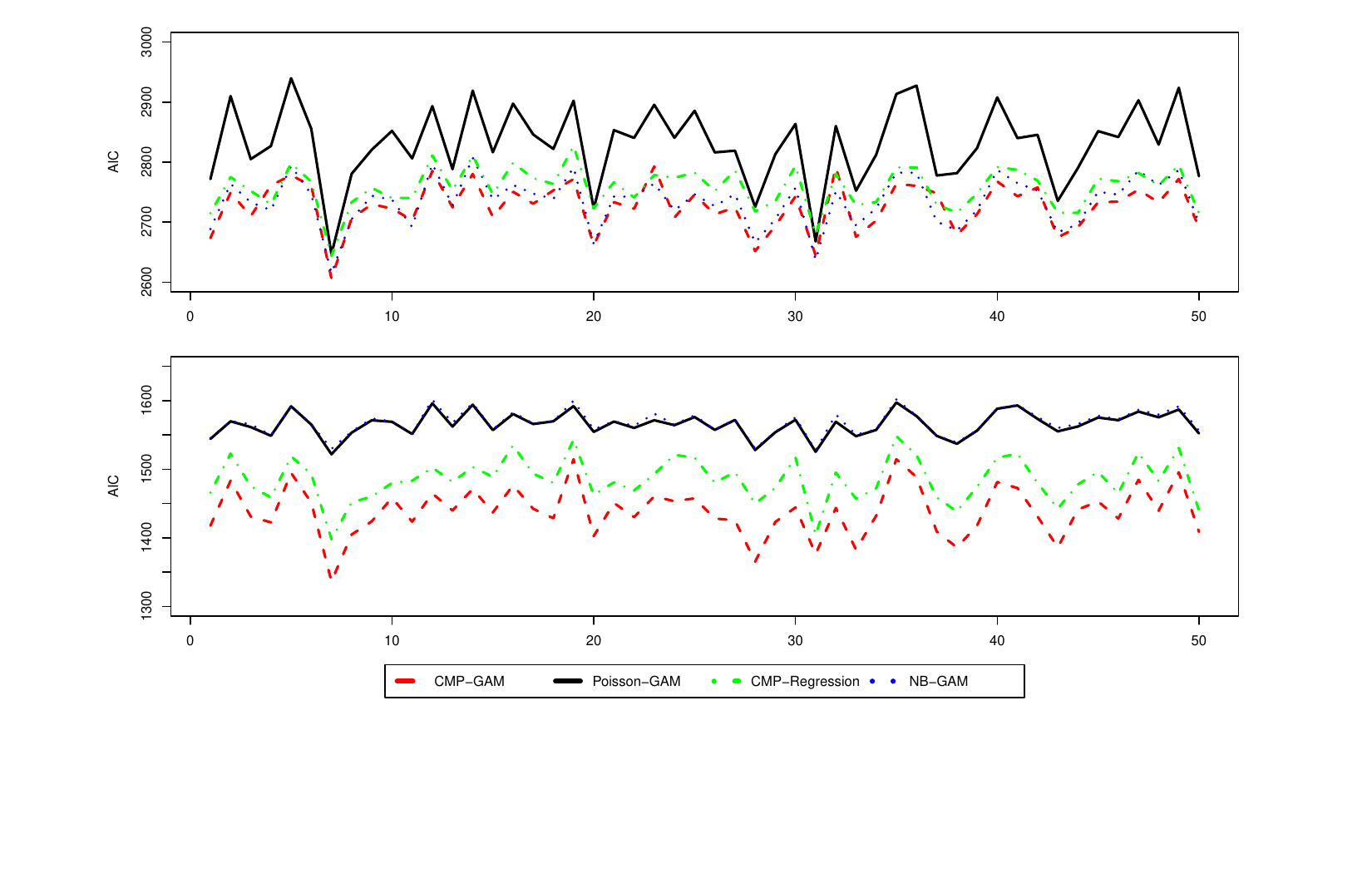}
	    \caption{Comparison of the $AIC$ values from CMP GAM, Poisson GAM, NB GAM and CMP Regression for the 50 bootstrap replications. The X-axis denotes the bootstrap sample. Top chart: over dispersion $\nu=0.5$; Bottom chart: under dispersion $\nu=2.5$.}
	    \label{fig:aic-gam}
	\end{figure}

	\begin{figure}
	    \centering
	    \begin{minipage}{\textwidth}
	    \includegraphics[height=0.3\textheight,width=0.9\textwidth]{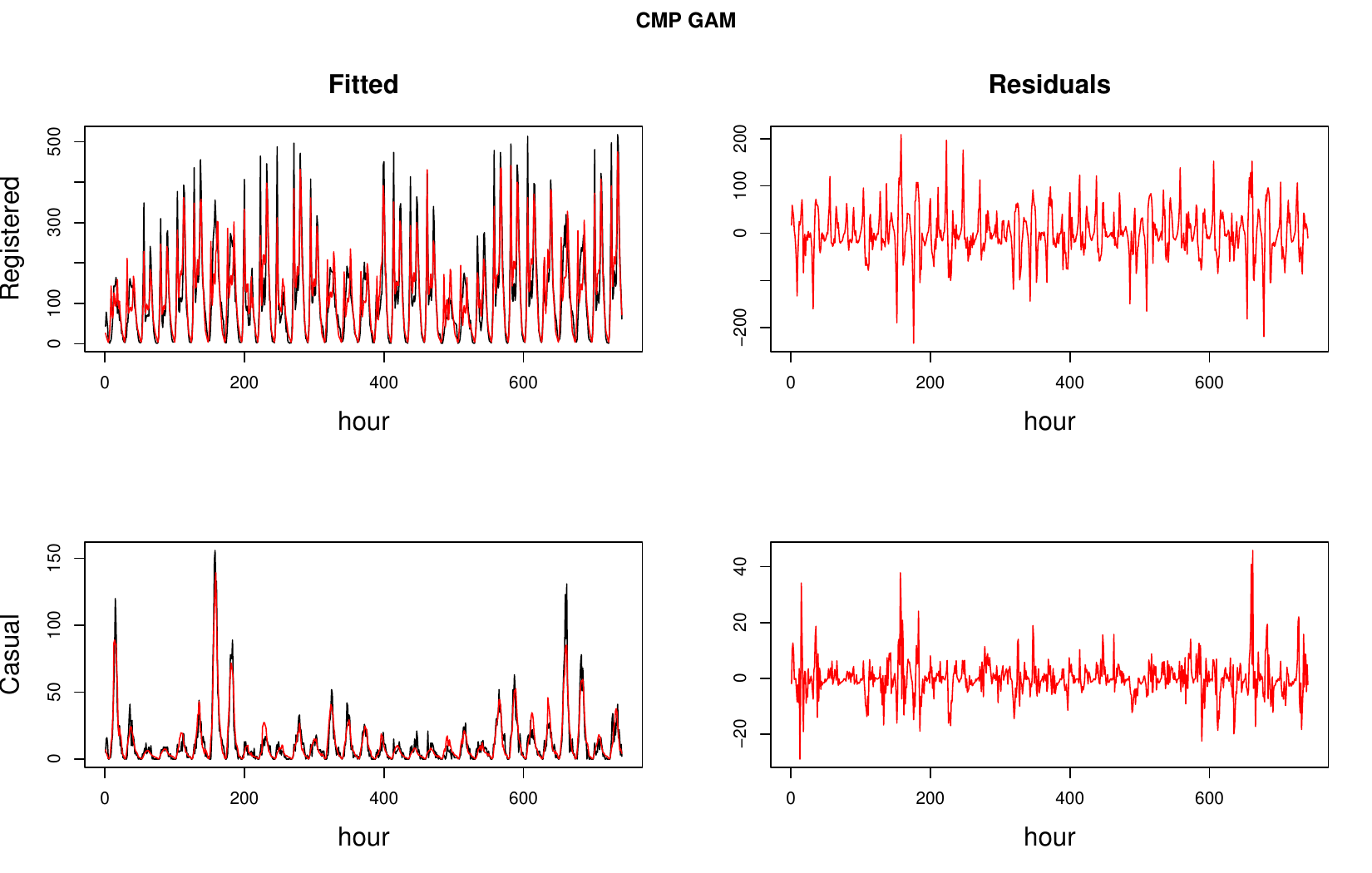}
	    \end{minipage}
	    %\vfill
	    \begin{minipage}{\textwidth}
	    \includegraphics[height=0.3\textheight,width=0.9\textwidth]{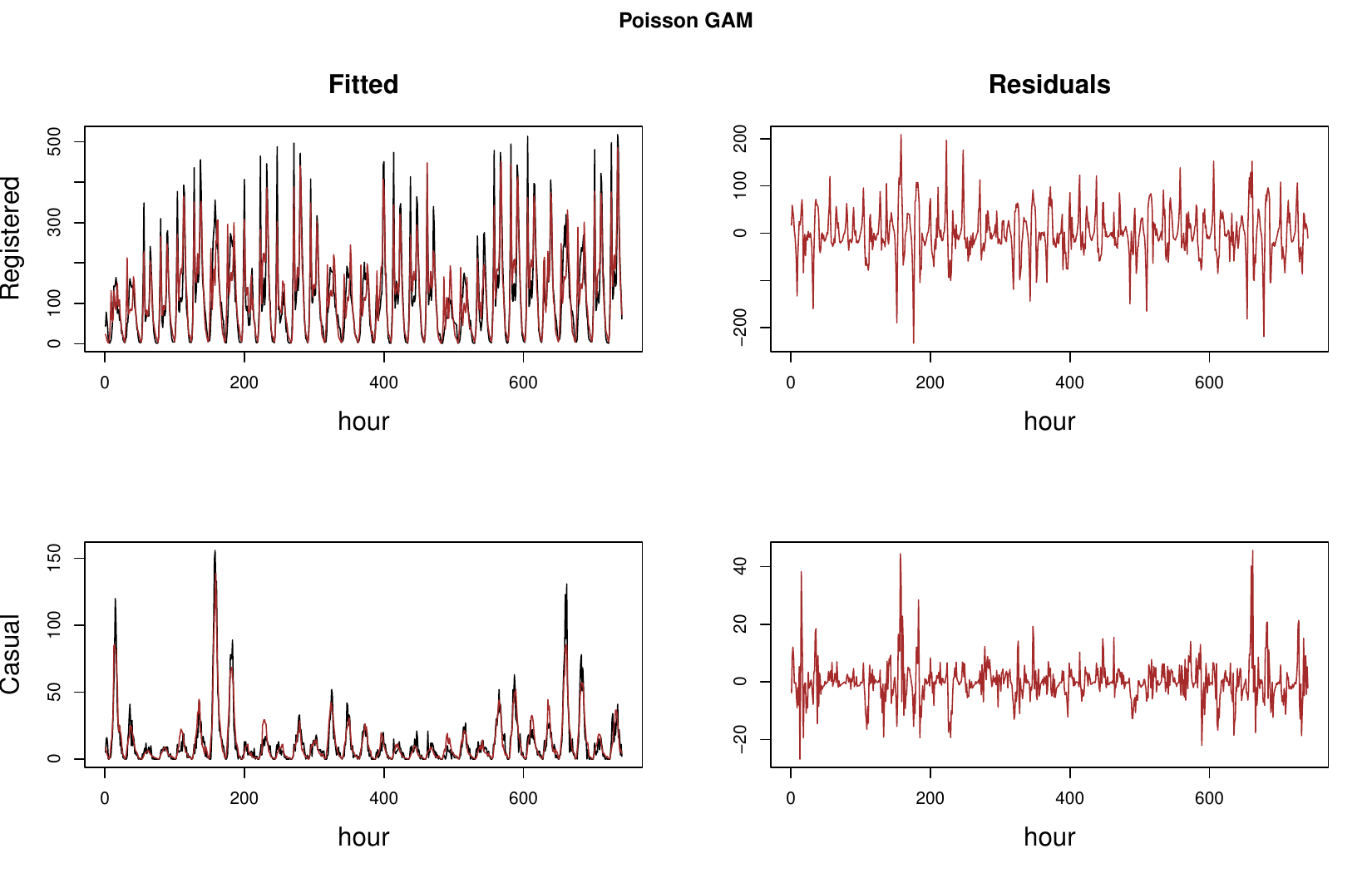}
	    \end{minipage}
	    %\vfill
	    \begin{minipage}{\textwidth}
	    \includegraphics[height=0.3\textheight,width=0.9\textwidth]{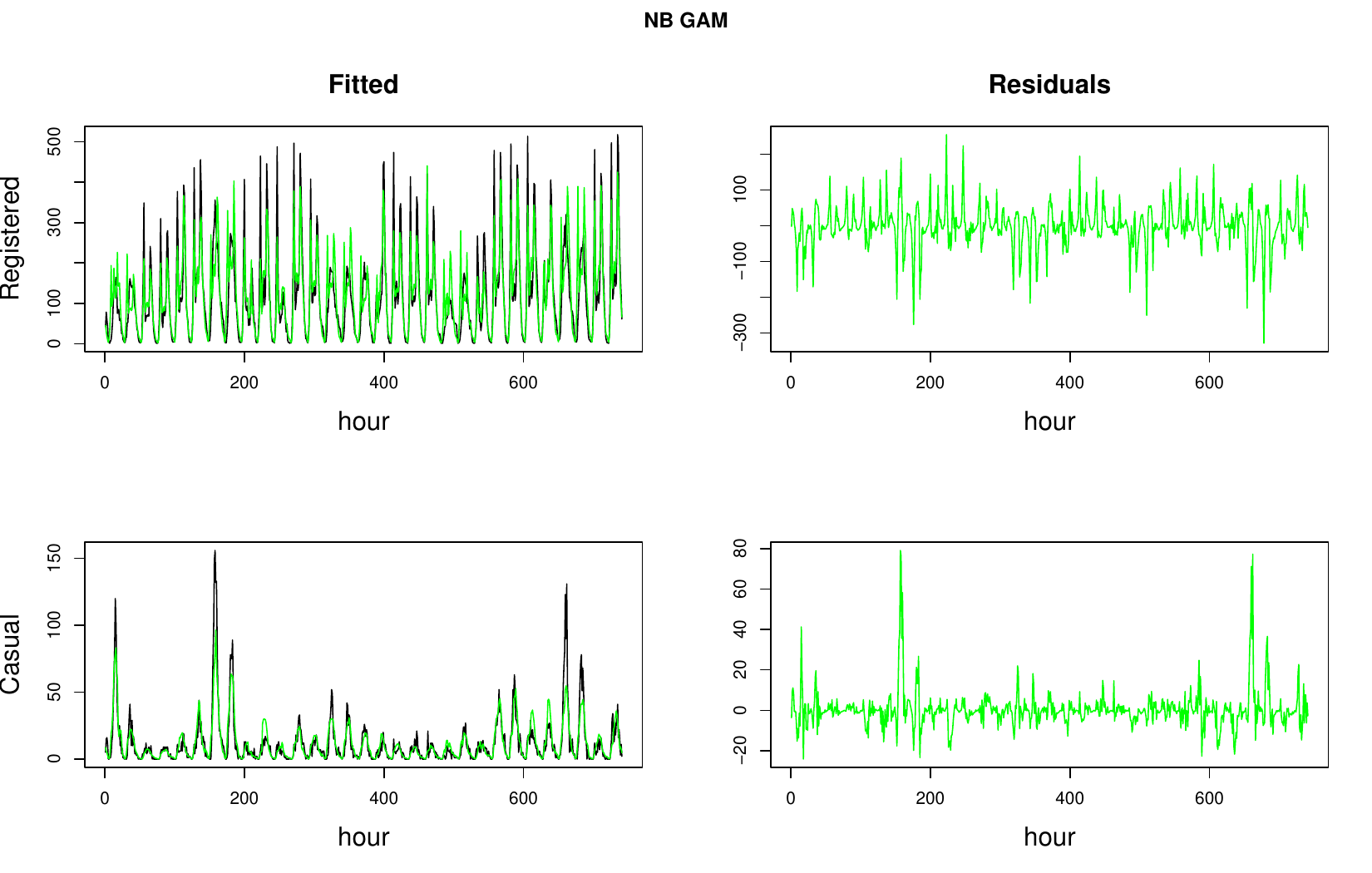}
	    \end{minipage}
	  \caption{Comparison of the fitted values and residuals from CMP and Poisson additive models with actual values. %The top plot is for registered users and the bottom plot is for casual users.
	  ({\color{black} \textemdash  True values}, {\color{red} \textemdash CMP GAM} , {\color{violet} \textemdash Poisson GAM} and {\color{green} \textemdash NB GAM}).}
		\label{fig:bikeshar-jan-pred}
	\end{figure}

\end{document}